\documentclass[10pt,journal]{IEEEtran}

\usepackage{amsmath,amssymb,amsfonts}
\usepackage{algorithmic}
\usepackage{textcomp}
\usepackage{xcolor}
\usepackage{multirow}
\usepackage[normalem]{ulem}
\useunder{\uline}{\ul}{}

\ifCLASSOPTIONcompsoc
  \usepackage[nocompress]{cite}
\else
  \usepackage{cite}
\fi

\usepackage[pdftex]{graphicx}
\graphicspath{{../figures/}}
\DeclareGraphicsExtensions{.pdf,.jpeg,.png,.jpg}
\usepackage[]{algorithm,algorithmic}
\usepackage{amsmath,amssymb,amsfonts}
\usepackage{algorithmic}

\usepackage{array}

\usepackage{url}

\usepackage{textcomp}
\usepackage{xcolor}
\usepackage{multirow}
\usepackage[capitalise]{cleveref}
\usepackage{flushend}
\usepackage{caption}
\usepackage{subcaption}
\newcommand{\red}[1]{#1}

\begin{document}
\bstctlcite{IEEEexample:BSTcontrol}

\title {SCV-GNN: Sparse Compressed Vector-based Graph Neural Network Aggregation}

\author{Nanda~K.~Unnikrishnan*, \IEEEmembership{Graduate Student Member, IEEE}; Joe~Gould*, \IEEEmembership{Student Member, IEEE}; 
and Keshab~K.~Parhi, \IEEEmembership{Fellow, IEEE}%
\IEEEcompsocitemizethanks{%
\IEEEcompsocthanksitem{N. K. Unnikrishnan, J. Gould, and K. K Parhi are with the Department of Electrical and Computer engineering at the University of Minnesota, Minneapolis, MN 55455 USA (e-mail: unnik005@umn.edu, gould146@umn.edu,  parhi@umn.edu).}%
\IEEEcompsocthanksitem{This work has been supported in part by the National Science Foundation under Grants CCF-1914749.}%
}%
}

\IEEEtitleabstractindextext{%

\begin{abstract}
Graph neural networks (GNNs) have emerged as a powerful tool to process graph-based data in fields like communication networks, molecular interactions, chemistry, social networks, and neuroscience. GNNs are characterized by the ultra-sparse nature of their adjacency matrix that necessitates the development of dedicated hardware beyond general-purpose sparse matrix multipliers. While there has been extensive research on designing dedicated hardware accelerators for GNNs, few have extensively explored the impact of the sparse storage format on the efficiency of the GNN accelerators. This paper proposes SCV-GNN with the novel sparse compressed vectors (SCV) format optimized for the aggregation operation. We use Z-Morton ordering to derive a data-locality-based computation ordering and partitioning scheme. The paper also presents how the proposed SCV-GNN is scalable on a vector processing system. Experimental results over various datasets show that the proposed method achieves a geometric mean speedup of $7.96\times$ and $7.04\times$ over CSC and CSR aggregation operations, respectively.  The proposed method also reduces the memory traffic by a factor of $3.29\times$ and $4.37\times$ over compressed sparse column (CSC) and compressed sparse row (CSR), respectively. Thus, the proposed novel aggregation format reduces the latency and memory access for GNN inference.

\begin{IEEEkeywords}
Neural Network Inference, Accelerator Architectures, Graph neural networks, Aggregation.
\end{IEEEkeywords}

\end{abstract}

}

\maketitle
\makeatletter{\renewcommand*{\@makefnmark}{}
\footnotetext{* (equal contribution)}\makeatother}


\IEEEdisplaynontitleabstractindextext


\section{Introduction}

Deep neural networks (DNNs) are brain-inspired models and have permeated into everyday facets of our lives~\cite{DNN_history}. These models have shown significant promise in the domains of image recognition~\cite{alexnet,VGG,resnet,googlenet}, speech~\cite{speech}, language~\cite{BERT,roberta} and medical diagnosis~\cite{medical}. Recently there has been a keen interest in exploring applications where the data is highly structured in the form of graphs~\cite{GCN,GAT,GIN,graphsage,gcn_survey} using graph neural networks (GNNs). This has wide-ranging applications from the performance of communication networks~\cite{routenet}, molecular interactions in chemistry~\cite{MPNN}, human relations through social media networks~\cite{social_media}, and brain function and disease analysis in neuroscience~\cite{medical_GNN}.
GNNs have shown significant promise as they are able to exploit the dependencies encoded in the graph structure across multiple layers, allowing for an effective relational inductive bias from the neural network design. 

There are several challenges introduced by graph computing. First, as graph sizes continue to grow exponentially, storing the data within the local memory becomes prohibitive. This necessitates further research into efficient methods to perform inference to enable further adoption. Secondly, GNN computations are highly memory and communication-intensive, requiring irregular memory access patterns and leading to high memory latency. 
Lastly, GNN adjacency matrices have a high degree of nonuniform sparsity ($\geq 99.9\%$), where most nodes contain very few edges and a few nodes contain the majority of edges. This results in poor data locality and heavy imbalances in the processing element workloads. This makes GNN workloads unsuited for general purpose processors~\cite{NvidiaWP_amp,TPUanalysis,TPUv3}, layer pipelining~\cite{Pipedream,Gpipe,layerpipe}, DNN accelerators~\cite{permdnn,eyerissv2_19,EIE,InterSCH}, and sparse matrix multiplication accelerators~\cite{sigma,fpga_sparse,3dic_sparse,extensor}.  The unique challenges above of GNNs have led to a plethora of solutions in both software~\cite{DGL,PyG}, accelerator~\cite{GenGNN,GNNerator,GNN_accel,versgnn,rubik,gnn_in_memory}, and HW/SW co-design~\cite{co_design} space. Most hardware solutions have largely focused on helping sparse accelerators mitigate the above challenges~\cite{GraphLily} rather than tailoring the solution to the characteristics of GNNs. This is inadequate to handle the irregular patterns of the aggregation operation. To mitigate the load balancing issue, GNN accelerators often employ some form of prepossessing of either the clever tiling strategies~\cite{zipper,HyGCN,Cambricon-G}, reordering of the nodes~\cite{GNNie,AWB-GCN,rubik}, or feature reuse~\cite{gnn_reuse,rubik}. While these preprocessing techniques can significantly help with workload balancing, they are impractical in real-time applications when each input graph is unique, and the preprocessing step is a recurring cost. The above hardware solutions also mitigate the load balancing issue by including complex queues, network-on-chips, and accumulators to distribute the workloads across PEs and collect the results. However, these changes bring significant overhead to the accelerator design, limiting its applications and generalizability.

Given these challenges, we propose a novel software-hardware co-design solution, \textit{sparse compressed vectors} (SCV), that optimizes the sparse format for maximizing hardware efficiency, and we develop an optimized hardware platform to exploit the proposed new format. The proposed format uses fixed-sized column vectors stored in a row-major format. The nature of storing the values within a block opposite of the order of the blocks themselves improves memory efficiency by balancing the input and output matrix priorities. 
From an architectural perspective, SCV has two improvements over standard sparse formats. First, the column-based vectors maximize the reuse of the input matrix for all the non-zeros in the array while allowing hazard-free parallelism. Furthermore, as the format implicitly stores the location of non-zero columns, it allows for efficient prefetching of the input matrix. Second, the row-major block processing order improves output matrix performance by accessing partial sums multiple times before evicting. The proposed block format allows exploiting existing cache blocking and data-locality strategies like Morton-Z-ordering~\cite{z-morton} (SCV-Z) to enhance memory efficiency further. The proposed data-locality-based ordering enables improved scalability with multiple processors through efficient partitioning. Given the new processing format, we design a queue-based general-purpose vector processor. The proposed architecture demonstrates SCV processing while also functioning as a general-purpose vector processor. 

The main contributions of the paper are as follows.
\begin{itemize}
    \item We propose SCV, a novel format, that significantly reduces the random access patterns during aggregation.
    \item We develop a novel processing order that prioritizes computation parallelism without the need for complex preprocessing or introducing workload imbalances.
    \item We introduce a blocking strategy that allows exploiting data locality strategies in the memory hierarchy, like Z-order, and improves the scalability of the design.
    \item We map the proposed format to a generalized queue-based vector processor that shows the simplicity in supporting the proposed format. 
\end{itemize}

The remainder of the paper is organized as follows. \cref{sec:background} focuses on the key GNN equations and computations and the benefits and limitations of existing sparse formats. \cref{sec:spase} describes the proposed SCV format and how it can be used for processing GNNs. \cref{sec:architecture} describes a general vector processor to support the proposed format. \cref{sec:evaluation} evaluates the proposed methodology. 
\cref{sec:related_work} describes the related work in the field. 
Finally, in \cref{sec:conclusion}, we summarize the paper's main conclusions.  
\begin{figure*}
    \centering
    \includegraphics[width=0.9\linewidth]{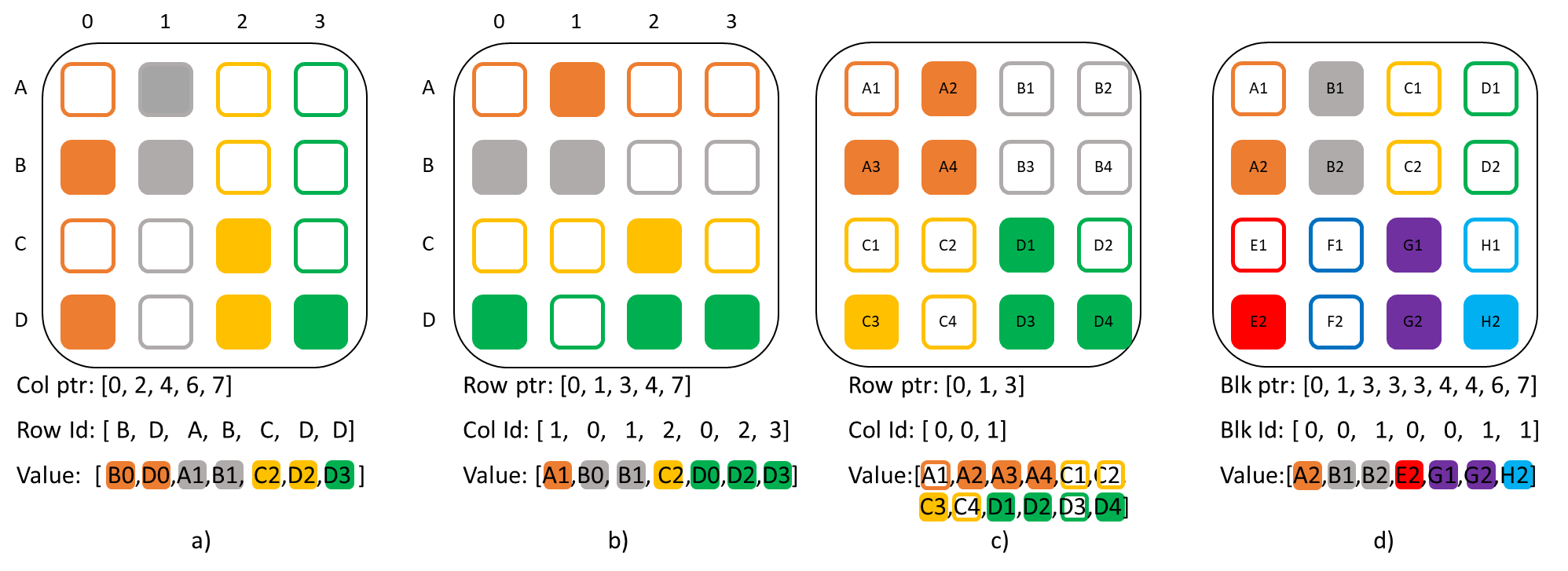}
    \caption{Various sparse representations with the pointer, index, and values arrays for a) CSC, b)  CSR,  c) BCSR, and d) SCV.}
    \label{fig:sparse_formats}
\end{figure*}

\section{Graph Neural Networks and Sparse Formats}
\label{sec:background}

\subsection{Graph Neural Networks}
Traditional neural networks like convolutional neural networks (CNNs) are designed to work with spatial or temporal data like images, speech, and videos~\cite{DNN_history,SzeTutotrial}. GNNs are a generalization of these neural networks to work on graph data structures that aim to collect features of each node from its K-hop neighbors. The success GNNs have had has led to the rise of various GNN architectures~\cite{GenGNN,GNNerator,GNN_accel,versgnn,GNNie,AWB-GCN,ENGN}. To effectively describe GNNs, we employ the general message passing paradigm~\cite{MPNN,DGL} as defined below. 

Consider the $h_v\in \mathcal{R}^{d_t}$, the feature for the node $v$ and $w_e \in \mathcal{R}^{d_e}$, the feature for an edge $e: u\rightarrow v$. The set of all feature vectors is defined as $\mathbf{H}^{(t)}$ for layer $t$. We can derive the message passing paradigm as follows~\cite{DGL}:

\begin{equation}
\begin{aligned}
    Edges: m_e^{(t+1)} &= \phi(h_v^{(t)}, h_u^{(t)}, w_e^{(t)}), (u, v, e) \in \mathcal{E} \\
    Nodes: h_v^{(t+1)} &= \psi(h_v^{(t)},\rho(\{m_e^{(t+1)}: (u, v, e) \in \mathcal{E}\}))
    \label{eq:MPNN}
\end{aligned}
\end{equation}

\noindent where $\mathcal{E}$ is the set of neighbor edges of node $v$, $\phi$ is the message function that combines incident edge and node features (\textit{combination function}), $\psi$ is an update function, and $\rho$ is the aggregating reduction function (\textit{aggregation function}). 
 
While node and edge-based definitions are beneficial, we can better understand the mapping of the above operations to hardware by defining the operations in terms of standard matrix/vector operations as shown below:

\begin{align}
    \mathbf{Z}^{(t)} &= \mathbf{H}^{(t)}\mathbf{W}^{(t)} \label{eq:sp_comb}\\
    \mathbf{H}^{(t+1)} &= \sigma(\mathbf{\hat{A}}\mathbf{Z}^{(t)})    
    \label{eq:sp_agg}
\end{align}

\noindent where $\mathbf{Z}^{(t)}$ is the combined feature matrix, $\mathbf{\hat{A}}$ is the weighted adjacency matrix for layer $t$, and $\sigma$ is a nonlinear operation that represents an activation function or pooling. \red{ \cref{eq:sp_comb} is a matrix version of the edge operation where the output $\mathbf{Z}^{(t)}$ maps to  $m_e^{(t+1)}$. \cref{eq:sp_agg} is the matrix version of $h_v^{(t+1)}$, where the messages from adjacent nodes are aggregated.} $\mathbf{\hat{A}}$ for special graph networks such GCN~\cite{GCN}, GraphSAGE~\cite{graphsage}, GIN~\cite{GIN}, and GAT~\cite{GAT} are described in these references respectively. 

\cref{eq:sp_comb} represents the combination operation between the previous layer's output and the weight matrix, and \cref{eq:sp_agg} represents the weighted aggregation step. Thus, defining GNNs in this form allows for the targeted development of accelerators for the combination and aggregation step.

\subsection{Sparse Formats}
The adjacency matrices used for aggregation are often ultra-sparse, with sparsities $\geq 99\%$. Additionally, the number of nodes in the graph is significant to the order of thousands to millions of nodes. The simplest way to store and process the sparse matrices is in the coordinate (COO) format, where we store each non-zero as a 3-element tuple of row index (row\_id), column index (col\_id), and its value. While the COO format is simple, it is not very efficient for storing or processing very large or ultra-sparse matrices. This warrants an exploration of sparse formats suited for aggregation. The aggregation step processes the output of the combination step,~\cref{eq:sp_comb}. We refer to the output of the combination step as the combined feature matrix, $\mathbf{Z}$. 

From \cref{eq:sp_agg}, we can analyze the memory access patterns for the aggregation step. Processing computations in a row of the adjacency matrix map to load the corresponding row of the output matrix. Similarly, processing elements in a column of the adjacency matrix map to a single row of the $\mathbf{Z}$ matrix. This can be used to analyze the effectiveness of the proposed sparse formats. We describe four baseline formats: compressed sparse column (CSC), compressed sparse row (CSR), block compressed sparse row (BSCR), and multipass (MP).

\subsubsection{Compressed Sparse Column (CSC)}
\begin{figure*}
    \centering
    \includegraphics[width=\linewidth]{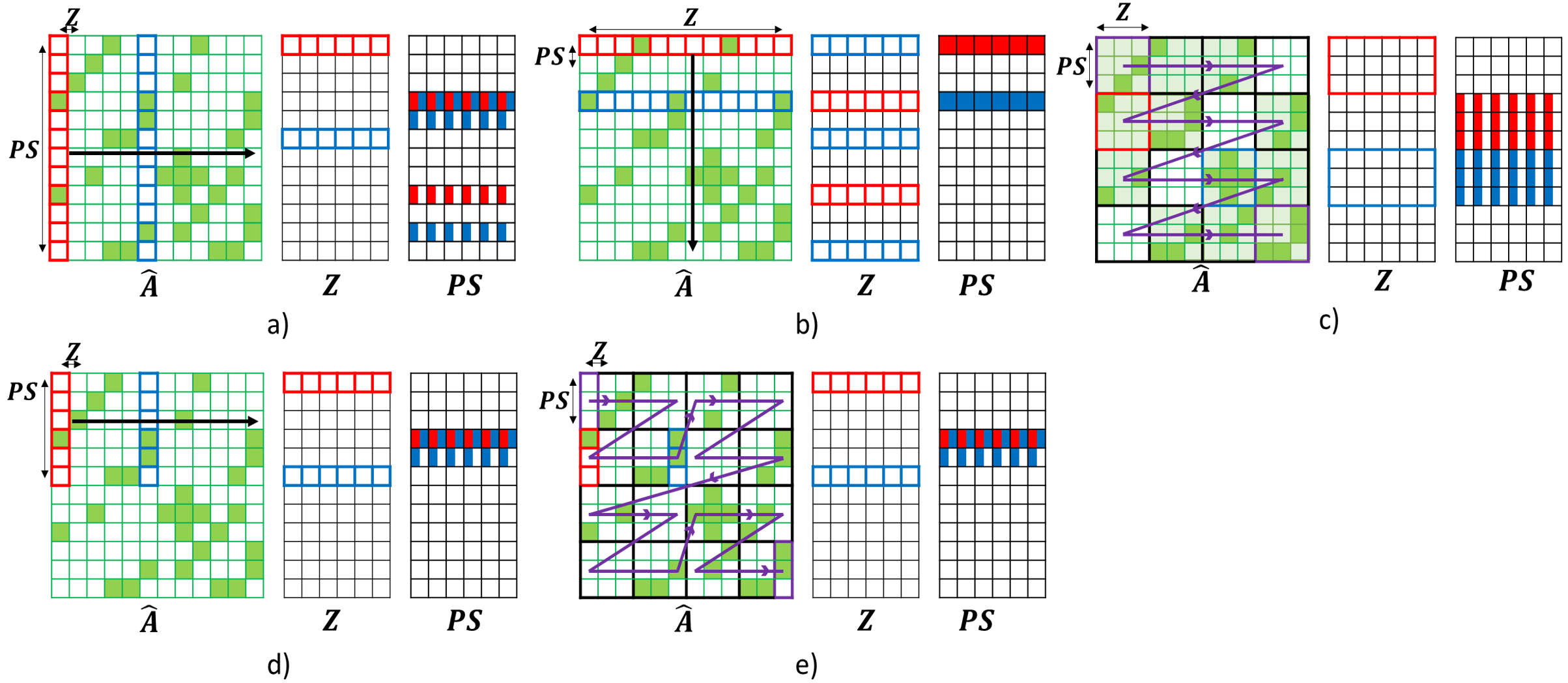}
    \caption{Comparisons of the matrix multiplication processing order on the adjacency matrix $\hat{\mathbf{A}}$ of the sparse formats: a) CSC, b) CSR, \red{c) BCSR}, d) SCV, and e) SCV with Z ordering (SCV-Z). \red{The blocks highlighted in red and blue represent two blocks within the adjacency matrix and the corresponding $\mathbf{Z}$ and $PS$ matrices rows loaded corresponding to the matrix multiplication computation.}}
    \label{fig:sparse_processing}
\end{figure*}

\cref{fig:sparse_formats}(a) shows the CSC representation of a sample sparse $4\times4$ matrix. The non-zeros are stored in an array in column-major format, the \textit{values} array. The corresponding row values for each non-zero are stored in the \textit{row id} array. The \textit{col ptr} array stores the starting location for each column in the values array \red{with the last entry pointing to the end of the values array. Each color represents one column  of the matrix.}  \cref{fig:sparse_processing}(a) shows the computation order for a CSC matrix multiplication operation. The operation iterates through each column of the adjacency matrix, loading all the non-zero elements. Each column also loads a single row of the $\mathbf{Z} $ matrix, and each non-zero loads the corresponding row of the output matrix's partial sums ($PS$). From an architectural perspective, CSC increases the reuse of the combined feature matrix, $\mathbf{Z}$, ensuring it is utilized before moving to the next row. However, this comes with irregular access to the $PS$ matrix. This is highlighted by the span of the $\mathbf{Z}$ and $PS$ shown in the top left of \cref{fig:sparse_processing}(a). 

\subsubsection{Compressed Sparse Row (CSR)}
\cref{fig:sparse_formats}(b) shows the CSR representation of a sample sparse $4\times4$ matrix. The non-zeros are stored in a linear array in row-major format, as shown by the \textit{values} array. The corresponding column values for each non-zero are stored in the \textit{col id} array. The \textit{row ptr} array stores the starting location for each row in the \textit{values} array \red{with the last value pointing to the end of the values array. Each color represents one row of the matrix}.  \cref{fig:sparse_processing}(b) shows the computation order for a CSR matrix multiplication operation. The operation iterates through each adjacency matrix row, loading all the non-zero elements. Each row also loads a single partial sum matrix row, and each non-zero loads the corresponding row of the $\mathbf{Z}$ matrix. From an architectural perspective, CSR increases the output matrix reuse, $PS$, ensuring it is computed before moving on to the next row. However, this comes at expense of irregular accesses of the combined feature matrix $\mathbf{Z}$. This is highlighted by the span of $\mathbf{Z}$ and $PS$ shown in the top left of \cref{fig:sparse_processing}(b).

\subsubsection{Block Compressed Sparse Row (BCSR)}
\cref{fig:sparse_formats}(c) shows the BCSR representation of a sample sparse $4\times4$ matrix. This is a blocked version of the CSR format, with each value replaced with a dense 2D block.  The non-zero blocks are each flattened in a row-major format. Each block is then stored contiguously in a linear array in a row-major format, as shown by the \textit{values} array. The corresponding column values for each non-zero block are stored in the \textit{col id} array. \red{This is analogous to the CSR array, where each entry tracks a block instead of a single value. For this example, there are two possible col ids, 0 and 1, as there are two columns of blocks}.  The \textit{row ptr} array stores the starting location for each non-zero block in the \textit{values} array. \red{This is analogous to the CSR array, where each entry points to the starting location of a row of blocks. For this example, the pointers are 0, 1, and 3, as one and two blocks are in the two rows, respectively. As each block is stored as a dense matrix, the two pointers only refer to the block locations rather than the individual values. Here each color represents one block of the matrix.}  The format trades off the benefits of reusing the $PS$ and $\mathbf{Z}$ matrices, allowing for regular access to both.  \red{The matrix multiplication operation iterates through each non-zero block of the adjacency matrix, as shown in \cref{fig:sparse_processing}(c). Each step loads the corresponding rows of the $\mathbf{Z}$ and partial sum matrices. One further advantage of this approach is that since each block is stored independently, we can exploit tiling and tile order to optimize memory efficiency. However, this comes at the cost of additional storage and memory access requirements, as non-zero blocks are always stored densely. For example, the block highlighted in red has a single value but is stored as a dense 3x3 requiring loading all of the corresponding rows of $Z$ and $PS$.}

\subsubsection{Multipass (MP)}

As seen with CSR, CSC, and BCSR, the order of operation can significantly impact the number of memory accesses. One approach to eliminate the influence of access order is to perform a memory-centric approach. The aggregation step can be seen as a scatter-gather operation between different nodes, thus, we can exploit existing enhancements for such workloads. A multiple-pass approach, or Multipass~\cite{multipass1,multipass2}, iterates through the matrix multiple times, only performing the aggregation if all the dependent variables are already loaded into memory or cache. This significantly reduces the number of misses in the cache and allows for a more regular access pattern to the DRAM. 
The data loaded into the cache is kept until it is completely processed or is evicted based on some predetermined thresholds. This approach trades off memory access regularity for increased computation workload. Specifically, the matrix multiplication must complete multiple rounds or passes over the input data until all the nodes have been processed. Furthermore, as data can be used for multiple rounds, intermediate results must be held locally for longer.

\section{Sparse Compressed Vectors}
\label{sec:spase}

\subsection{Improvements to Existing Sparse Formats}
The baseline formats were designed with general sparse matrix multiplication in mind. However, improvements can still be tailored toward the ultra-sparse models of graph neural networks. The goal of the proposed method is to take advantage of the BCSR format while reducing its liabilities. The proposed method targets three specific improvements.
\begin{itemize}
    \item Reducing the overhead of the format by storing the internal blocks as sparse. 
    \item Exploring format options to maximize GNN computational efficiency.
    \item Exploiting sparsity for efficient memory accesses.
\end{itemize}

First, the major shortcoming of BCSR is the dense manner in which it stores its internal blocks. One possible solution is the use of compressed sparse blocks~\cite{CSB} format, which modifies the internal blocks to be stored in a sparse format. The CSB format stores the information in fixed size square blocks, usually a power of 2. The non-zero blocks are each flattened in a row-major format. Each block is stored contiguously in a linear \textit{values} array. The corresponding row and column values for each non-zero are stored in the \textit{row id} and \textit{col id} arrays. The main difference between this and a COO format is that the row id and col id store the relative addresses with respect to the block and not the entire matrix. Thus this requires $\log_2 B$ bits, where  $B$ is the block size. This is significantly lower than the $\log_2 N$ bits required in the COO format as $B<<N$. The block pointer array, \textit{blk ptr}, stores the starting location for each non-zero block in the values array. The user can determine the block order based on the application. The format incorporates the benefits of BCSR while reducing the memory requirement.  The operation iterates through each row of non-zero blocks in the adjacency matrix. Each step loads the corresponding rows of the $\mathbf{Z}$ and partial sum matrices. It also exploits tiling and tile order to optimize memory efficiency. This significantly improves the adjacency matrix performance but has little impact on the $\mathbf{Z}$ and partial sum matrices. 

Second, to improve computational GNN efficiency, we modify the operation order within the blocks themselves. CSB stores the non-zero values contiguously but does not give preference to columns or rows in their internal COO format. We propose using a column-major storage format that leads to a two-fold improvement. Column-based processing allows for parallelizing the computations within the block without creating imbalances. If the column is sufficiently large enough, multiple rows in the partial sum array can be processed in parallel without conflict. Also, column-based processing allows for regular accesses to the $\mathbf{Z}$ matrix, as it adopts a sequential row-by-row access pattern.

Third, we explore how to prioritize reads from memory to improve memory access efficiency. For example, if an entire row within a sparse adjacency block is zero, we need not load the corresponding partial sum row. Similarly, a zero column in the adjacency block implies that the corresponding row of the $\mathbf{Z}$ is not required. A row-centric ordering would negate the earlier parallelizing benefits and is suboptimal.  Therefore to maximize efficiency, rather than working directly with square tiles like CSB, we propose further dividing the block into column vectors, and test this in \cref{sec:evaluation}. Thus, we introduce the proposed \textit{Sparse Compressed Vector} (SCV) format. 

\subsection{Sparse Compressed Vectors (SCV)}
\cref{fig:sparse_formats}(d) shows the SCV representation of a sample sparse $4\times4$ matrix. The matrix is divided into a series of fixed-size block column vectors (2 in this example). The vector's contents are stored continuously, and each non-zero column vector is stored in a row-major format, as shown by the \textit{values} array. The corresponding location values within a vector for each non-zero are stored in the \textit{blk id} array. \red{In this example, the blk id associated with a value can be either 0 or 1 depending on its location within the column vector.} The \textit{blk ptr} array stores the starting location for each vector in the values array. \red{Each column vector in the matrix has a corresponding  blk ptr, 8 column vectors of size two are present in this example. Here each color represents one column vector of the matrix.} The SCV format can be interpreted as using the CSB format with a block width of 1 column. \cref{fig:sparse_processing}(d) shows the computation order for an SCV matrix multiplication operation. The matrix multiplication operation iterates through each vector of the adjacency matrix, loading all the non-zero elements. Each vector also loads a single row of the $\mathbf{Z}$ matrix. $PS$ matrix rows are loaded corresponding to the rows of the block vector. From an architectural perspective, there are two improvements from using SCV. First, the column-based vectors maximize the $\mathbf{Z}$ matrix reuse for all non-zeros in the array. Furthermore, as the format implicitly stores non-zero columns locations, which allows for prefetching the $\mathbf{Z}$ matrix efficiently. Second, the row-major processing order improves $PS$ matrix performance as the fetched $PS$ rows are reused multiple times before being evicted. This is highlighted by the span of the $\mathbf{Z}$ and $PS$ shown in the top left of \cref{fig:sparse_processing}(d).

It is important to distinguish SCV from tiled CSC and CSR operations. Tiling CSC or CSR merely breaks existing rows or columns into tiles without changing the processing order of the computation. While these changes allow for improving parallelization, they do not improve memory access efficiency. Specifically, tiled CSC and CSR are still inefficient with respect to the $PS$ and $\mathbf{Z}$ matrices, respectively. SCV changes  the processing order of the computation such that it processes vectors orthogonal to its storage order. These changes retain the benefits of a tiled CSC operation while reducing the inefficiency with the $PS$ matrix.

\subsection{Z-Ordering for Improved Cache Efficiency (SCV-Z)}

Thus far, we have not addressed the system's optimal ordering of SCV blocks. Though we initially describe SCV with a row-major block order, in principle, SCV can support any user-specific order based on the application. We explore this possibility with a modified Z-Morton ordering~\cite{z-morton}.  Z-Morton is a storage format that can map multidimensional data to a single dimension while prioritizing the locality of the data. It is a recursive storage format that first stores all elements in the top-left quadrant, then the top-right, bottom-left, and bottom-right quadrants. The same layout is used recursively within each quadrant. There were multiple choices for choosing the operation order beyond Z-Morton, such as U-Morton and Hilbert layouts. We chose the Z-Morton layout due to its simplicity of encoding and limited preprocessing.
However, as SCV uses vectors instead of square tiles, we use a modified version of Z-Morton ordering that considers a set of column vectors as a single block. For simplicity, we choose the set size as the number of rows of the column vector.  \cref{fig:sparse_processing}(e) shows a sample processing of matrix multiplication operation with SCV and a fixed block-sized Z-order (SCV-Z). The processing is broken up recursively into graph tiles. These tiles are then processed using the SCV format in the Z-Morton order preserving the locality of the different tiles in the memory hierarchy. Though we only show a 2-dimensional Z-order tiling of the adjacency matrix, this can easily be extended to a 3-dimensional Z-order by including the tiling of the combined feature matrix.

The proposed approach allows for easy processor scalability by virtue of its processing order. Using the processing line marked in purple in \cref{fig:sparse_processing}(e), we can arrive at a new storing order for the \textit{blk ptr}, \textit{blk id}, and \textit{values} arrays that are shown in \cref{fig:sparse_formats}(d). Thus, blocks between different processors can be mapped evenly, efficiently distributing the non-zeros of the matrix when stored in the new order. This has two advantages over baseline architectures. First, any subsequence from the processing order also preserves data locality, and second, the smaller block sizes allow for fine-grain partitioning compared to row or column-based partitioning. The proposed format can be easily statically generated from the COO format and is nearly equivalent to creating a CSR or CSC matrix.

\section{Architecture}
\label{sec:architecture}

\begin{figure*}
    \centering
    \includegraphics[width=\linewidth]{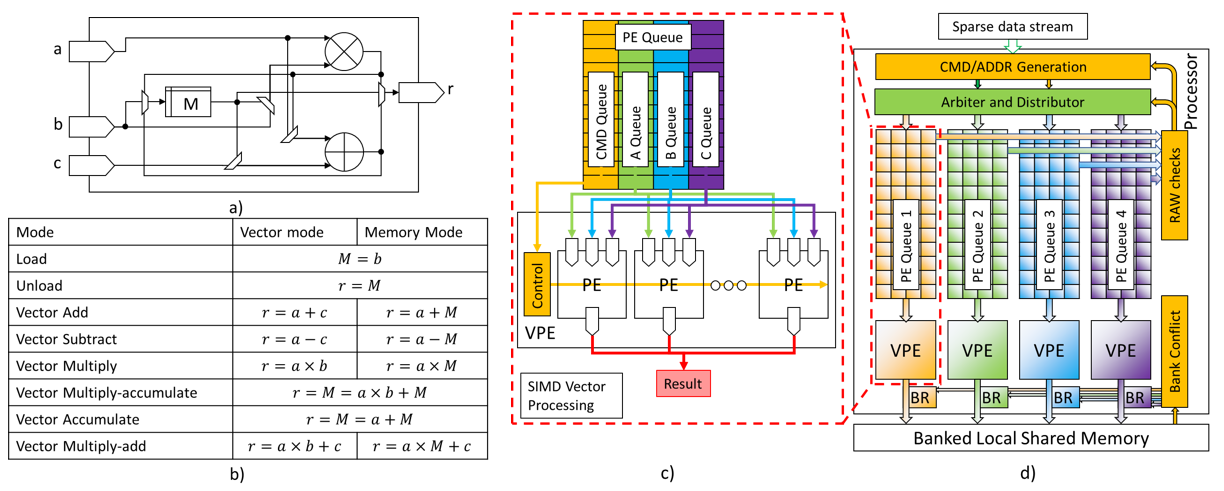}
    \caption{Architecture of the proposed queue-based vector processor. a) Structure of a single floating-point processing element. b) Supported set of operations for the PE. c) A vector processing element consisting of as many PEs as the width of the vector and its corresponding input queue. d) Complete vector processor architecture consisting of multiple VPEs, local memory, and control logic. }
    \label{fig:arch}
\end{figure*}

\subsection{Graph Processor}
We design and develop a multi-purpose configurable \textit{processing element} (PE) that can support various floating-point operations. \cref{fig:arch} (a) shows the overall architecture of the proposed PE. The PE consists of three input ports ($a$, $b$, and $c$), an output port ($r$), a local floating-point register ($M$), five multiplexers, a floating-point adder, and a floating-point multiplier. The PE has two modes of operation: a \textit{vector mode} that only uses the input and output ports, and a \textit{memory mode} that uses the internal memory register. The list of supported floating-point operations for the PE is shown in the table in \cref{fig:arch} (b). The PE is purposely designed for general-purpose processing as a fast parallel \textit{vector processor engine} (VPE) for single instruction multiple data (SIMD) operations. The proposed VPE is shown in \cref{fig:arch} (c). The VPE accepts a single command that is broadcast to all of its PEs. Each PE accepts three floating-point values from the respective queues for its three input ports. The number of PEs, $N_{PE}$ in a VPE, determines the size of the vector for SIMD processing. This allows for the VPE to support a variety of instructions beyond the traditional operations required for aggregation and combination. Subsequent sections of the architecture will highlight how these operations support the required functionality. Finally, the overall architecture of the \textit{graph processor} is shown in \cref{fig:arch} (d). Each processor consists of $N_{VPE}$ VPEs, each with its own queue for instructions and data. 

The processor uses multiple controllers for the data flow's different aspects. First, the \textit{cmd/addr generation} block takes the input data stream and performs a command and address translation. The input commands are mapped to the PE commands shown in \cref{fig:arch} (b) and map the input data to the local address within the \textit{banked local shared memory}. Second, The arbiter and distributor distribute the workload among the different VPEs based on availability and hazards. Last, the memory controller interfaces with the local memory to monitor bank or memory conflicts and dynamically stalls the processor. 

\red{While we do propose a new accelerator, care was taken to ensure that the multi-level processor would be as simple as possible, with the main improvements and novelty coming from the proposed SCV-Z format. This was done to ensure that the proposed format could work efficiently in a general vector processor to maximize its utility.}

\begin{figure}
    \centering
    \includegraphics[width=\linewidth]{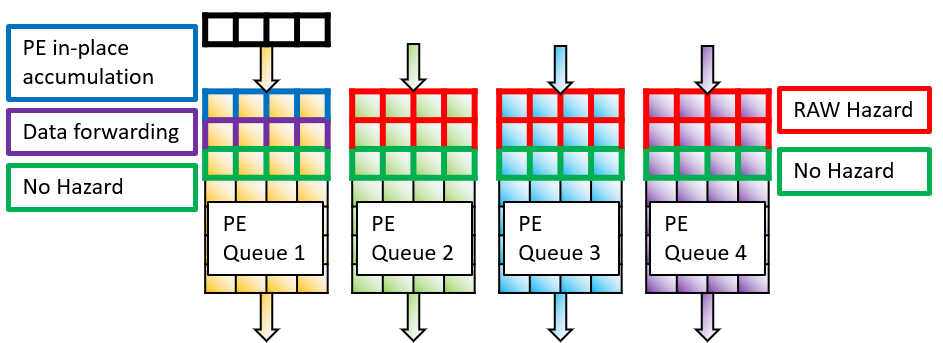}
    \caption{Potential RAW hazards for the input to PE queue 1 and their mitigation strategy in the proposed architecture. The red boxes highlight cross-queue RAW hazards. The blue and purple boxes highlight potential hazards within the same queue and how they can be mitigated. The green boxes highlight the location for which there are no hazards and beyond. }
    \label{fig:hazards}
\end{figure}

\subsection{Queues: Distribution and Hazard Handling}

The design uses data queues to create an asynchronous interface to the data stream. This helps reduce stalls by creating a buffer against data conflicts in the design. \cref{fig:arch} (c) shows a close-up of one sample queue architecture. Each queue consists of four internal queues, a command queue containing SIMD, and three data queues corresponding to the three inputs of the VPE. The command queue has a depth of $D$ and $4$ bits width to support all instruction types. Each data queue has a depth $D$ and width of $Wd_{addr}$ bits, where $Wd_{addr}$ is the width of the local address containing the required floating-point data. Each data queue can operate in two modes by loading either a scalar value that is broadcast to all $N_{PE}$ PEs of the VPE or a sequential set of $N_{PE}$ values corresponding to a vector. The data that is loaded from memory is mapped to the corresponding PE as shown in \cref{fig:arch} (c). The queues are designed as asynchronous FIFOs to allow for easier data flow control without the need to introduce stalls. 

As each queue handles data from different addresses, it is important to properly account for hazards. The memory model assumes that there is a two-cycle latency from when a result is written to a memory location and when it can be read back by any VPE. As such, locations accessed at least three cycles apart do not cause hazards, and in \cref{fig:hazards}, are highlighted as green outlined boxes. 
We look at the nature of the aggregation and combination operations to analyze hazards. At its core, all operations can be written in the form of $\mathcal{C}=\mathcal{A}\times\mathcal{B}+\mathcal{C}$, where $\mathcal{A}$ and $\mathcal{B}$ are read-only matrices. Therefore the architecture must only account for \textit{read-after-write} (RAW) hazards if the output $\mathcal{C}$ is required for subsequent computations. We handle these hazards with the following modifications. 
First, RAW hazards occur if the output address, $\mathcal{C}$, conflicts with the output address of any of the other parallel queues. \cref{fig:hazards} shows this as highlighted in red outlined boxes. To avoid this type of hazard, we perform a RAW hazard check and ensure that data mapped to the same output address are mapped to the same VPE or stalled. Second, within the same queue, there are three scenarios: the conflict is one cycle away (blue outlined boxes on PE Queue 1), the conflict is two cycles away (purple outlined boxes on PE Queue 1), or the conflict is three cycles away (green outlined boxes on PE Queue 1). For the one-cycle scenario, we handle the hazard by replacing the initial \textit{Vector-Multiply-Add} VPE command with a \textit{Vector-Multiply-Accumulate} command. This allows for the result to be stored for the immediate next cycle. In the two-cycle scenario, we implement data forwarding within the PE as an output buffer. When this scenario is detected, the data is forwarded to the input of the VPE, bypassing the memory block. Last, data three cycles or more away from a conflict create no hazard.

The arbiter and distributor block assigns the incoming data stream into the respective queues. The block first resolves cross-queue RAW hazards by assigning conflicting data to the same queue. The block then resolves within queue hazards with alterations to the command. The arbiter is designed to operate at a higher throughput than the queues and PE so that it can always keep the queues full. 

\red{ The above arbitration scheme was designed to not place any undue burden on the vector processor. Possible ways to improve the RAW hazard handling include dynamic workload distribution between VPEs and support for cross-VPE accumulation and reduction. Also, it is worth noting that with the proposed system, stalls are reduced significantly, and additional improvements may not be significant. }

\subsection{Banked Local Shared Memory}

The proposed dataflow requires access to high bandwidth local memory as each cycle, we require $3 \times N_{VPE}$ data reads and $N_{VPE}$ writes, where $N_{VPE}$ is the number of VPEs within a processor. Within the dataflow also, there are differing requirements for each data type in the computation $\mathcal{C}=\mathcal{A} \times \mathcal{B} + \mathcal{C}$. The inputs $\mathcal{A}$ and $\mathcal{B}$ require only the ability to perform $N_{VPE}$ parallel reads each, and $\mathcal{C}$ requires $N_{VPE}$ parallel read and writes. We use four enhancements to the local memory system to support these requirements.
First, we segregate the data within the local system into dedicated memories for $\mathcal{A}$, $\mathcal{B}$, and $\mathcal{C}$, allowing for tailored access based on the requirement. 
Second, the memories are designed to support a limited broadcast capability if there are parallel reads with the same address. This is useful when the weighted adjacency matrix ($\mathbf{\hat{A}}$) or the combined feature matrix ($\mathbf{Z}$) is reused over multiple computations. 
Third, we make use of four-port SRAM modules \cite{synopsys-multiport,renesas-4-port} to increase simultaneous read and write accesses. These allow for four parallel reads or writes access to the memory. The $\mathcal{A}$ and $\mathcal{B}$ matrices would configure the memory to allow four parallel reads each. Similarly, the $\mathcal{C}$ matrix would be configured for two reads and two writes.
Fourth, when the $N_{VPE}$ is greater than the data memory bandwidth, we employ memory banking~\cite{multi-port-mem-fpga}. Banked memories divide memory locations among multiple SRAMs (banks), allowing each additional bank to provide an additional two reads and two write ports for the $\mathcal{C}$ matrix or four read ports for the $\mathcal{A}$ and $\mathcal{B}$ matrices. However, if the number of requests exceeds a bank's bandwidth, this could cause a bank conflict. The memory processor resolves bank conflicts by stalling the requesting VPE. 

\begin{figure}
    \centering
    \includegraphics[width=\linewidth]{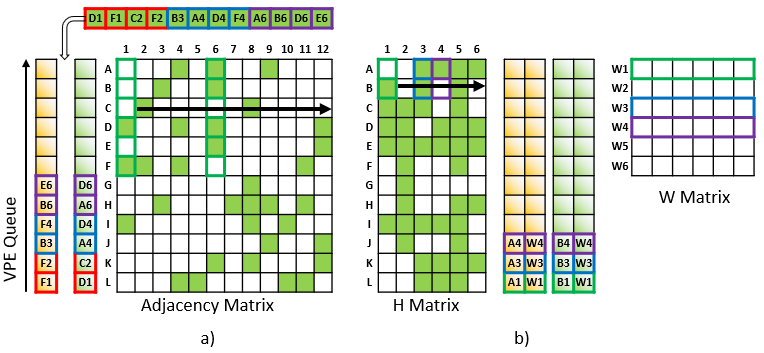}
    \caption{Mapping the aggregation and combination operations to multiple VPEs. a) The aggregation operation iterates through the list of non-zeros and assigns them greedily to the queues after appropriate hazard checks. b) The combination operation iterates through the non-zero column vectors and assigns them with the corresponding weight row the the queue. }
    \label{fig:order}
\end{figure}

\subsection{Aggregation Operation}

For the aggregation operation, we treat it as a sparse matrix multiplication between the ultra-sparse adjacency matrix $\mathbf{\hat{A}}$ and the dense combined feature matrix $\mathbf{Z}$ as shown in \cref{eq:sp_agg}.
In a direct aggregation case, as in the case of GCN, the rows of the feature matrix are added together based on the presence of ones in row of $\mathbf{\hat{A}}$. Other GNN models can be considered special cases of weighted aggregation where the ones of the adjacency matrix are replaced with appropriate weights, the degree of the nodes for GCNs, or the attention values for GATs. \cref{fig:order} (a) shows the processing order for the weighted adjacency matrix while performing aggregation. In this example, the number of VPEs is 2 and the size of the column vector is 6. The arbiter fills non-zero data into the queues while prioritizing mitigating RAW hazards. For each non-zero, the address of the rows of the combined feature matrix, $\mathbf{Z}$ (corresponding to the column of the adjacency matrix) and address of the output partial sums matrix, $PS$ (corresponding to the row of the adjacency matrix), are loaded into the queues. The resulting partial sums are written back into the same shared memory location. The $PS$ matrix is loaded into the shared memory once at the beginning and only ejected when moving to a new set of rows. The rows of the $\mathbf{Z}$ can easily be prefetched by observing the set of non-zero blocks in the SCV format. Similarly, as the data is stored in the queue in increasing order of the $\mathbf{Z}$ rows, a row may be evicted as soon as it  is no longer addressed in the queues. When processing from the queues, the VPE reads the address for the $PS$, $\mathbf{\hat{A}}$, and $\mathbf{Z}$ and loads the values from the shared memory. The results are written back to the same $PS$ address. This process is repeated until all the elements of the weighted adjacency matrix have been processed. 

\subsection{Combination Operation}

Though the proposed format does not directly improve the combination operation, the proposed architecture is general enough to support combination. We treat them as a sparse matrix multiplication between the sparse feature matrix, $\mathbf{H}$, and the dense weight matrix, $\mathbf{W}$ to create the combined feature matrix $\mathbf{Z}$ as shown in~\cref{eq:sp_comb}. \cref{fig:order} (b) shows the processing order for the $\mathbf{H}$ matrix while performing combination. 
The processor is first pre-loaded with the partial sums, $PS$ of $\mathbf{Z}$ of size $N_{VPE}\times N_{PE}$.  The processor performs an output stationary matrix multiplication operation where the row of $\mathbf{W}$ is broadcast to each VPE, and each scalar in the column vector of $\mathbf{H}$ is sent to one VPE and internally broadcast to all PEs. The addresses of the rows and column vectors are loaded into the queues for processing. The $PS$ matrix is loaded into the shared memory once at the beginning and only ejected when moving to a new output block. The module can skip a computation if all the elements of the $\mathbf{H}$ vector are zero. 
The rows of the $\mathbf{W}$ can be prefetched sequentially as we process the columns of the $\mathbf{H}$ matrix. The $\mathbf{W}$ matrix can be evicted once it is completely consumed in the queues.  When processing from the queues, the VPE reads the addresses for the $\mathbf{H}$ and $\mathbf{W}$ and loads the values from the shared memory. This process is repeated until all the output blocks have been processed. The proposed architecture is not limited to the above data flow and can easily be modified to support a weight or input stationary data flow.

\section{Evaluation}
\label{sec:evaluation}

\begin{figure}[t]
    \centering
    \includegraphics[width=0.9\linewidth]{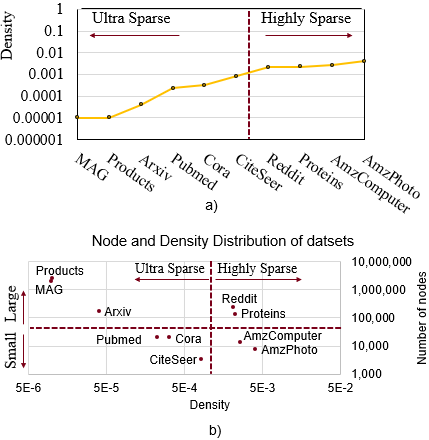}
    \caption{Characteristic of the datasets used for evaluation. a) The datasets are sorted by their sparsity. The datasets are divided into ultra-sparse and highly-sparse based on their characteristics during evaluation. b) Size of the graph versus the sparsity of the graph. The datasets are further divided into large and small graphs. }
    \label{fig:density-vs-dataset}
\end{figure}

\subsection{Methodology}

We evaluate the advantages of the proposed method for GCNs applied to various small and large datasets. The datasets vary in node density as shown in \cref{fig:density-vs-dataset}. Each experiment on these datasets varies the number of neurons in each layer of the GCN and the number of processing elements. \red{We sweep the configuration and hyperparameters taken directly from a number of GNN models, such as GCN~\cite{GCN}, GraphSAGE~\cite{graphsage}, GIN~\cite{GIN}, and GAT~\cite{GAT}, to generate aggregated results.} We developed a simulation tool to model the computational aspects for aggregation and combination to evaluate the proposed method. The tool calculates statistics such as the number of cycles and on-chip SRAM memory access for inference.
The tool supports both traditional dataflows as well as the proposed dataflow. We evaluate the proposed method and compare it with three baselines: a compressed sparse column (CSC) approach, a compressed sparse row approach (CSR), and a multiple pass (MP) approach.
The tool also models the shared memory system, generating a memory access trace. We use this memory access trace to evaluate the cache and DRAM performance with Ramulator~\cite{ramulator}. Ramulator is configured with default HBM settings, 1Gb/s bandwidth.

For evaluation of the performance of SCV-GNN, we used the benchmark graph datasets listed in \cref{fig:density-vs-dataset} a). The datasets are distributed by density, categorizing them into ultra-sparse and highly-sparse datasets. \cref{fig:density-vs-dataset} b) characterizes the sparsity with respect to the size of the graph, further categorizing them as large and small graphs. 
The datasets are summarized in \cref{tab:dataset-table}. \red{Note that the ogbn prefix is omitted for space in subsequent references.}

\begin{table}[t]
\caption{Characteristics of the datasets used for evaluation.}
\label{tab:dataset-table}
\resizebox{\columnwidth}{!}{%
\begin{tabular}{
p{0.21\linewidth}|r|r|>{\raggedleft\arraybackslash}p{0.07\linewidth}|>{\raggedleft\arraybackslash}p{0.12\linewidth}}
\hline
Dataset             & Nodes   & Edges     & Feature size & Adjacency density$\%$ \\ \hline
ogbn-mag            & 1939743 & 21111007  & 128          & 5.61E-06          \\
ogbn-products       & 2449029 & 61859140  & 100          & 1.03E-05          \\
ogbn-arxiv          & 169343  & 1166243   & 128          & 4.07E-05          \\
Pubmed              & 19717   & 88651     & 500          & 2.28E-04          \\
Cora                & 19793   & 126842    & 8710         & 3.24E-04          \\
Citeseer            & 3327    & 9228      & 3703         & 8.34E-04          \\
Reddit              & 232965  & 114615892 & 602          & 2.11E-03          \\
ogbn-proteins       & 132534  & 39561252  & 8    & 2.25E-03          \\
Amazon CoBuy Computer & 13752   & 491722    & 767          & 2.60E-03          \\
Amazon CoBuy Photo    & 7650    & 238163    & 745          & 4.07E-03          \\ \hline
\end{tabular}%
}
\end{table}

The tool is designed to model the proposed data flow as follows. 
The data for aggregation and combination is streamed into the processor as shown in \cref{sec:architecture}.
This is processed cycle-wise to determine the number of multiply and accumulate operations (MACs), internal register reads/writes, inter-PE communication and on-chip SRAM reads/writes from the array. All comparisons to baselines are iso-MAC and iso-memory to ensure a fair comparison. The results report memory accesses and memory access savings pertain to both SRAM and DRAM, as the proposed method is designed to target data locality within both their access patterns. The reuse of variables from the proposed sparse format ensures maximum utilization at all levels of hierarchy.

We model the memory hierarchy in two steps. First, the local shared memory bank is modeled within the simulator, keeping track of all the variables loaded into the processor. Second, if the data is not present in the shared local memory or needs to be written out, the simulator generates a memory trace file of all access going to and from the lower memory levels. We use this trace file in Ramulator to model a complete memory hierarchy to test the lower levels of memory. The local shared memory is modeled as partitioned between 64kB for the adjacency matrix, 64kB for the combined feature matrix, and 256kB for the output matrix for a processor memory of 384kB. Ramulator models the shared memory, a 2MB cache, and the DRAM with 4GB.
We simulate our architecture with 8 VPEs of 64 PEs each, for 512 floating-point MACs in total. We give the baselines we simulate against an identical MAC configuration. Our architecture's VPE queues use a depth of 16. We choose an SCV vector size of 512 for our evaluation unless otherwise noted. Datasets missing from the results are due to the memory limitations of the simulation hardware and software available. However, the proposed architecture is still expected to run aggregation in these cases.

We only present results with respect to aggregation, as only the adjacency matrix can be assumed to have significant sparsity for general GNN processing. However, for combination, the proposed architecture attains latency that is at least as good as systolic arrays~\cite{kungsys}, which are often used for the dense combination step \cite{HyGCN}. This is because our VPE architecture generally supports a higher communication bandwidth with shared memory, eliminating systolic array warm-up and cool-down times that cause underutilization. 

\begin{figure}[t]
    \centering
    \includegraphics[width=\linewidth]{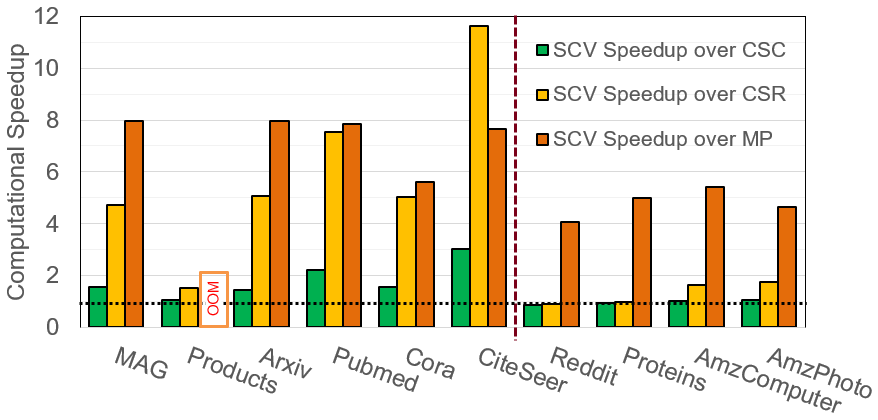}
    \caption{Speedup in computations cycles normalized to other formats without memory-induced stalls. Ultra-sparse datasets show higher speedups with the proposed method compared to highly-sparse datasets.}
    \label{fig:speedup-computations}
\end{figure}

\subsection{Computation Cycles}
We explore the effect the various sparse formats have on the number of computation cycles. The results do not include latency from memory which is discussed later. \cref{fig:speedup-computations} shows the comparative analysis of SCV-GNN versus traditional sparse formats in terms of computational cycles. All the results show the relative speedup of SCV over CSC, CSR, and MP. Column-based processing, such as CSC and SCV, leads to significant improvements in the number of computation cycles, especially over row-based approaches such as CSR. 
As shown in the figure for the ultra-sparse datasets, SCV-GNN has a significant speedup over CSR with a geometric mean of $5.03\times$ speedup. The speedup in the highly-sparse datasets is lower, with a geometric mean speedup of $26\%$. Compared to CSC, the proposed method has a geometric mean speedup of $36\%$.

\begin{figure}[t]
    \centering
    \includegraphics[width=\linewidth]{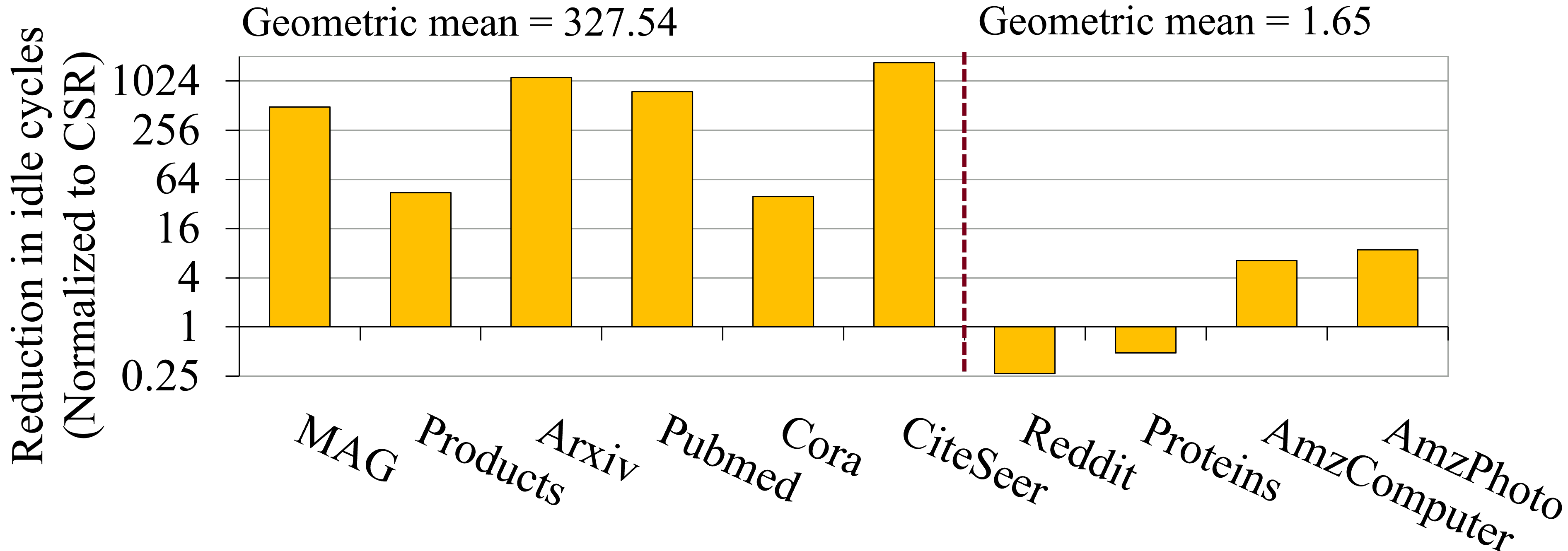}
    \caption{Reduction in the number of idle cycles normalized to CSR. Ultra-sparse datasets show higher reduction when compared to highly-sparse datasets. }
    \label{fig:idle-cycles}
\end{figure}

There are several reasons why there is an improvement in computational performance. First, column-based approaches maximize parallelism allowing different VPEs to work in parallel on different output rows and for large intervals before re-addressing the same output row. Second, the CSC and CSR approaches map a fixed set of output rows to a PE, which limits performance and leads to significant workload imbalances resulting in idle cycles.
The significant speedups over CSR can be attributed to the reduction in idle cycles, as shown in \cref{fig:idle-cycles}. The proposed method achieves a geometric mean of $327\times$ reduction in the ultra-sparse datasets and a $1.65\times$ reduction in highly-sparse datasets. This also explains the difference in the performance of the ultra-sparse and highly-sparse datasets. 

\begin{figure}[t]
    \centering
    \includegraphics[width=\linewidth]{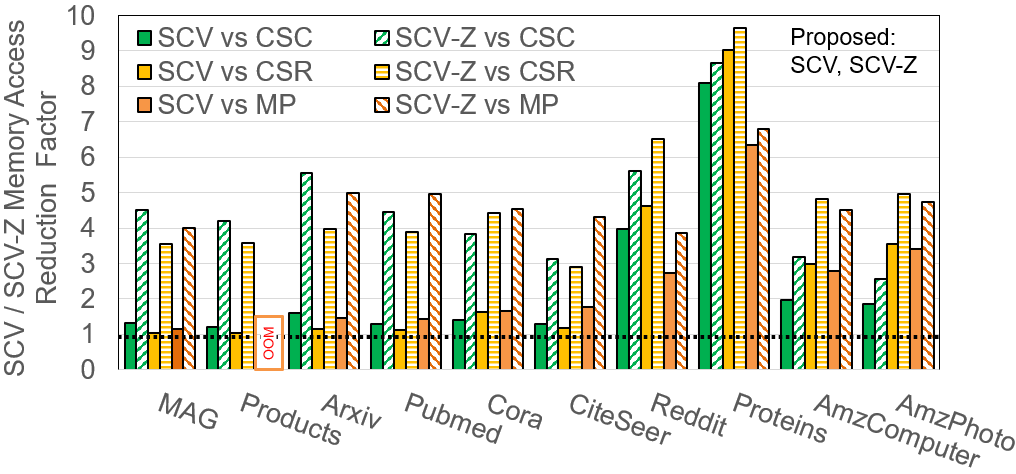}
    \caption{Reduction in the overall memory traffic to the cache. Each column shows the improvement factor of the proposed SCV/SCV-Z over baseline sparse processing formats. SCV outperforms the baselines for all test cases.}
    \label{fig:memory-access}
\end{figure}

\subsection{Memory Access}

\cref{fig:memory-access} shows the comparative analysis of SCV-GNN versus traditional sparse formats in terms of memory access from the processor.  All the results show the relative reduction in the number of memory accesses of SCV/SCV-Z over CSC, CSR, and MP. The results are normalized by dividing the total memory traffic of the baseline divided by the total memory traffic of SCV/SCV-Z to show the improvement factor.

As shown in the figure, for the highly-sparse datasets, SCV-Z has a significant reduction over CSR and CSC with a geometric mean of $4.37\times$ and  $3.29\times$ improvement, respectively. The method improve over the ultra-sparse datasets with CSR and CSC to a geometric mean reduction in memory accesses of $13\%$ and $34\%$, respectively. 
This can be attributed to the better utilization of the data within the processor due to the column-wise processing as well as the limited partial sum range. The Z-order for SCV allows for efficient memory management, significantly reducing the number of data accesses.

\begin{figure}[t]
    \centering
    \includegraphics[width=0.9\linewidth]{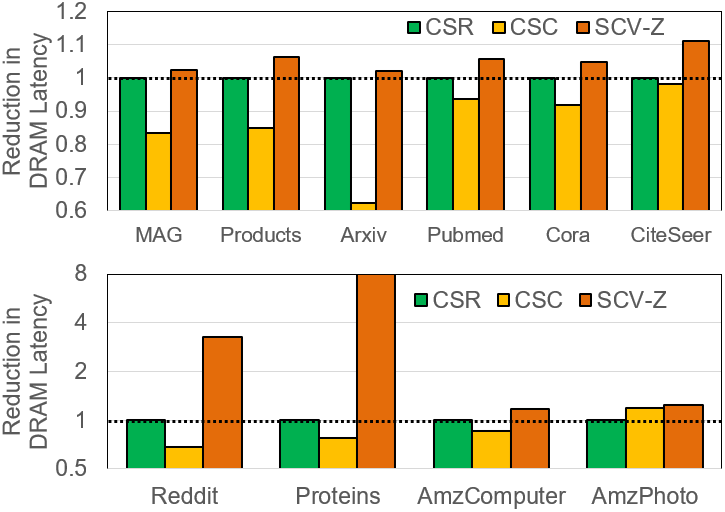}
    \caption{Reduction in the mean access time (MAT). Each column is normalized to show the improvement over CSR (MAT of CSR / MAT of selected format). The results are shown separately for the ultra-sparse (top) and highly-sparse (bottom) datasets. SCV outperforms the baselines for all test cases.}
    \label{fig:DRAM}
\end{figure}

\subsection{DRAM Mean Access Time}
\cref{fig:DRAM} shows the comparative analysis of SCV-GNN versus traditional sparse formats in terms of mean DRAM latency and the \textit{mean access time} (MAT). All the results show the relative reduction in the memory traffic to the DRAM of SCV-Z compared to CSC and CSR. MAT is measured as DRAM active cycles/number of requests from Ramulator. Each column is normalized to show the improvement over CSR (MAT of CSR/MAT of selected format).
The figure shows for the highly-sparse datasets, SCV-Z has a significant reduction with a geometric mean improvement of $2.48\times$ over CSR. The method does not significantly reduce the ultra-sparse datasets with a geometric mean of $4\%$ over CSR. Overall, SCV outperforms CSC by a factor of $24\%$ and $2.88\times$ for ultra-sparse and highly-sparse datasets, respectively. 
This is from better data utilization within the processor due to the column-wise processing and the limited partial sum range.  As we are only measuring the MAT, most of the memory benefits have been captured at the output of the shared local memory.

\begin{figure}[t]
    \centering
    \includegraphics[width=\linewidth]{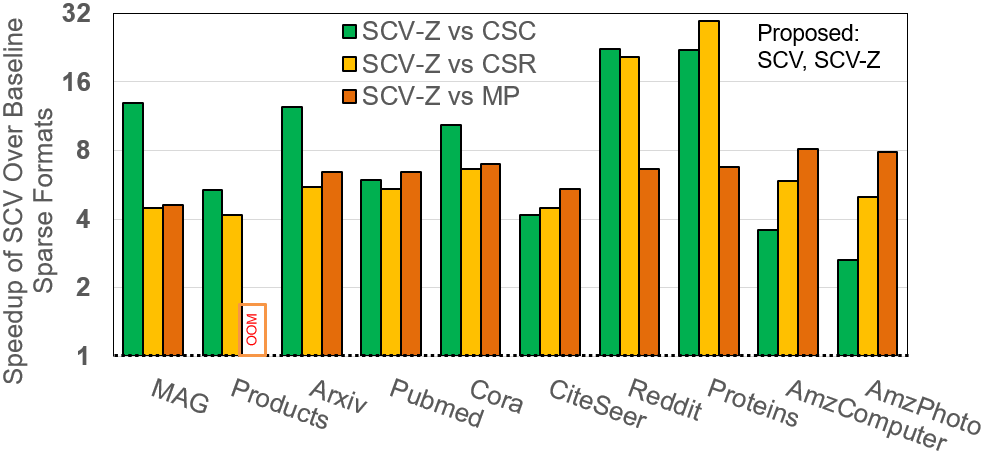}
    \caption{Overall speedup of the proposed SCV-Z method compared to other sparse formats, including memory-induced stall. The speedups are shown in a logarithmic scale.}
    \label{fig:aggr_res}
\end{figure}

\subsection{Overall Aggregation Performance}
Using Ramulator, we measure the mean access time(MAT) of the lower levels of memory, the cache, and the DRAM. \red{For our simulations, we have modeled a Cache and a DRAM as defined in the methodology. Ramulator takes the memory traces from our simulator and internally measures the CPU latency and MAT that is fed back into our simulator.} 
\red{This MAT reflects the average time the underlying memory subsystem takes to respond to compute engine for the given trace. We feed this MAT back into our simulator to accurately estimate the overall performance of the proposed methods and the baselines. When the simulator is measuring the time to  access data in the local scratchpad, it can be retrieved in a single cycle if it is a hit (data is present in the scratchpad). However, it if is a miss, we use MAT to estimate the average time to retrieve the data from the memory subsystem. On a miss, the corresponding VPE is considered stalled for the duration.}
Thus, we summarize the overall performance improvement for the aggregation in~\cref{fig:aggr_res}. The figure shows that the proposed SCV-Z format significantly outperforms all sparse formats for all datasets achieving a geometric mean speedup of $7.96\times$, $7.04\times$, and $6.51\times$ over CSC, CSR, and MP, respectively, with $10\times\sim29\times$ improvement on some of the larger datasets. 

\begin{figure}[t]
    \centering
    \includegraphics[width=\linewidth]{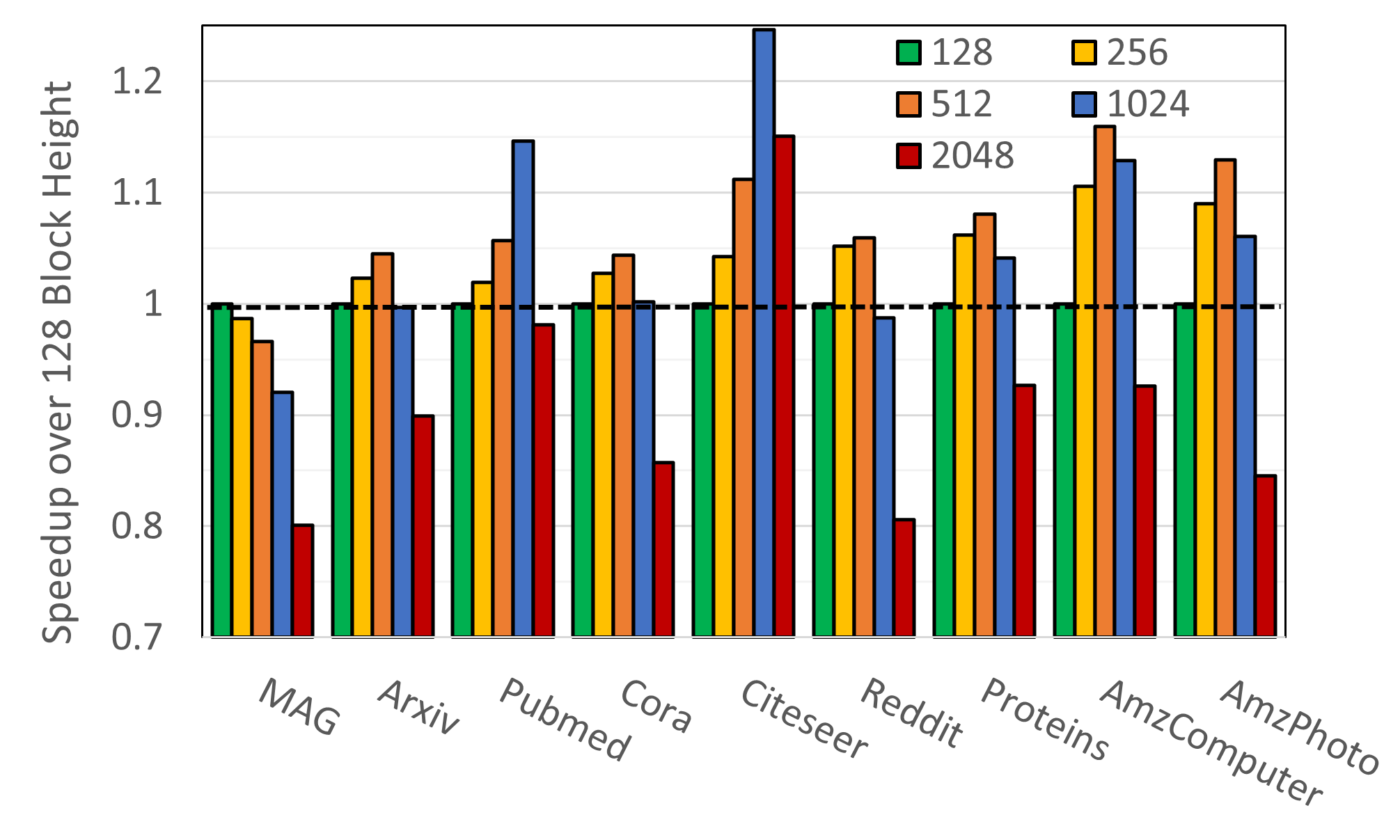}
    \caption{Speedup of various SCV vector heights compared to a height of 128.}
    \label{fig:scv_hgt_res}
\end{figure}
\begin{figure}[t]
    \centering
    \includegraphics[width=\linewidth]{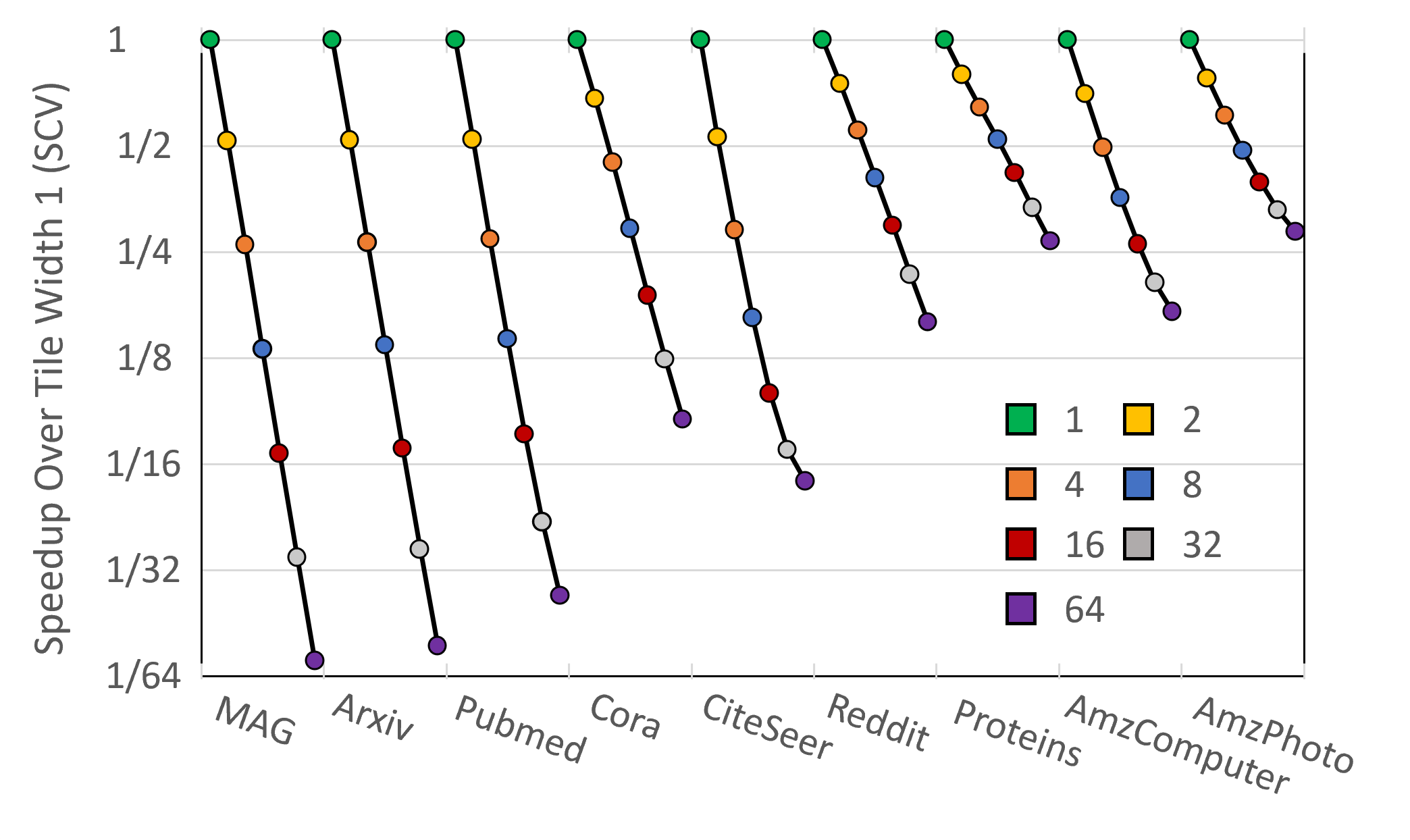}
    \caption{Speedup of SCV-like formats with multiple columns compared to SCV to a width of 1. Results are graphed using a logarithmic scale. The number of columns is swept from 1 to 64 in powers of 2. The results are normalized to SCV with a column width of 1. All formats use a column height of 64. }
    \label{fig:a_col_res}
\end{figure}

\subsection{SCV Parameter Sweep}
Thus far, we chose the block height of 512 based on the amount of data that would fit within the local memory. However, exploring how the choice of SCV vector height impacts total latency across the various datasets is of interest. To ensure a fair comparison, we fix the system's total memory and number of PEs while changing the number of rows in an SCV column. Increasing the height of the SCV column increases the span of the \textit{PS} matrix accessed, requiring more memory to store at once, but increases input feature reuse. To ensure an iso-memory comparison, the feature dimension of the partial sums stored is reduced to compensate for the increased span of \textit{PS}, keeping the total memory allocation unchanged. \cref{fig:scv_hgt_res} shows the effect on latency by varying the column height from $128$ to $2048$ in powers of 2. It is shown that for each dataset, an optimal height exists where the maximum overall reuse of the adjacency, input feature, and partial sum matrices is achieved. However, this point appears to be dependent on the sparsity and size of a given dataset, as there is no single optimal size for all datasets. From~\cref{fig:scv_hgt_res}, we show that $512$ and $1024$ are the most performant choices with a geometric mean speedup of 7.1\% and 5.5\% over $128$, respectively, validating our earlier results.

We further explore alternatives to the SCV format that does not limit tiling the adjacency matrix to single-column wide vectors. To explore this further, we investigate the impact on latency by sweeping the number of column vectors from $1$ to $64$ in powers of 2. Because increasing the column vectors only affects what data is accessed, varying it does not change the amount of on-chip memory. \cref{fig:a_col_res} summarizes the effect of tile width on the latency of the system. From the figure, it is shown that the performance of the system deteriorates as the number of columns in a tile increases.
This can be attributed to the more efficient reuse of SCV over CSB or similar multiple-column tiles. Increasing the columns in a tile increases the number of non-zero tiles while distributing the non-zeros across multiple columns. The issue lies in the fact that even if just a single non-zero is present within a tile, all the corresponding rows in the combined feature matrix must be accessed, even if there are no non-zeros within its corresponding column. This decreases the granularity in which zeros within the adjacency can be skipped. This affects denser datasets less, as SCV has fewer opportunities for zero skipping, but ultra-sparse datasets see a significant slowdown. 

\begin{figure}[t]
    \centering
    \includegraphics[width=\linewidth]{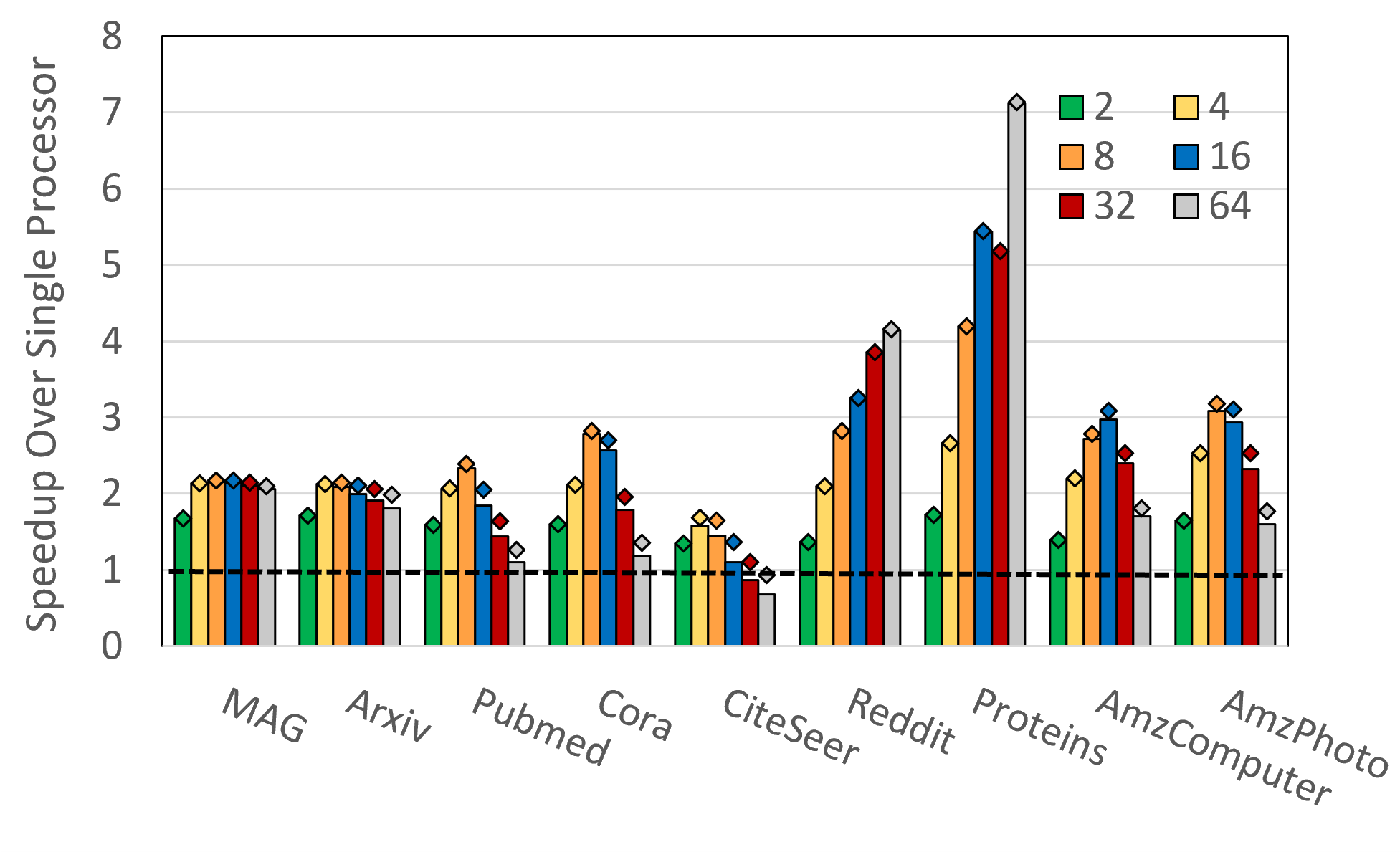}
    \caption{Speedup from increasing processors 2 to 64 in powers of 2. Diamond-shaped points represent the highest obtainable speedup if mergers of results from multiple processors weren't required, while bars represent the actual speedup.} 
    \label{fig:multiproc_res_overlay}
\end{figure}

\subsection{Scalability Analysis}
We also study the effect of increasing the number of assigned processors on overall latency. We scale the system by increasing the number of processors and their caches but keep the DRAM bandwidth fixed. We statically split the workload using the proposed Z access order of the adjacency, combined feature, and output matrices so that each processor handles roughly an equal number of adjacency non-zeros. The proposed SCV method allows for fine-grain partitioning at the vector level, unlike complete columns or rows in CSR or CSC approaches. Furthermore, the data-locality preserving nature of the Z-order ensures that the cache efficiency of the newly introduced caches does not significantly deteriorate.

As multiple processors can introduce data hazards, we use a controller to detect when two processors are working on the same output tile. When this occurs, the controller redirects the second processor to use a buffer region of the memory for a partial sum read and write to avoid a RAW hazard. At the end of processing the entire aggregation, multiple results for the same output are merged before continuing to the next task. This requires additional memory access and computation time overheads. There are no additional communication overheads as all inter-processor data transfer happens through the shared memory system. The simulator is modified to account for these overheads in the multi-processor scenario.

We sweep the number of processors from 2 to 64 in powers of 2, and the speedup shown is normalized to a single processor as shown in~\cref{fig:multiproc_res_overlay}. The points marked by diamonds show the absolute speedup from increasing the number of processors, and the bars show the speed-up after accounting for the overheads on latency from merging the results.

From~\cref{fig:multiproc_res_overlay}, it is shown that in ultra-sparse datasets, the speedup increases as we increase the number of processors up to 8 or 16 processors, after which adding more processors is detrimental. Further analysis shows that the latency is primarily limited by their memory access time. Initially, when the number of processors increase, the extra on-chip caches increase cache-to-processor bandwidth, reducing the average access time. However, it also decreases the reuse within each cache, as accesses to the same addresses become scattered across multiple processors. This becomes dominant with more than 16 processors, outweighing the benefit from the increased bandwidth. Denser datasets see significant reuse even with smaller chunks and still see significant speedup due to reduced computation time and increased cache bandwidth.

From~\cref{fig:multiproc_res_overlay}, we show that the actual speedup is fairly close to that without mergers. Increasing the number of processors increases the latency penalty as there are more opportunities for conflicts, requiring mitigation. However, these conflicts only occur when processors work on the same output at the same time. The chance of a conflict decreases with larger datasets, as the time between accessing the same output tile becomes longer. Thus the proposed method is able to achieve speedups near the maximum afforded by the memory system.

\begin{figure}[t]
    \centering
    \includegraphics[width=\linewidth]{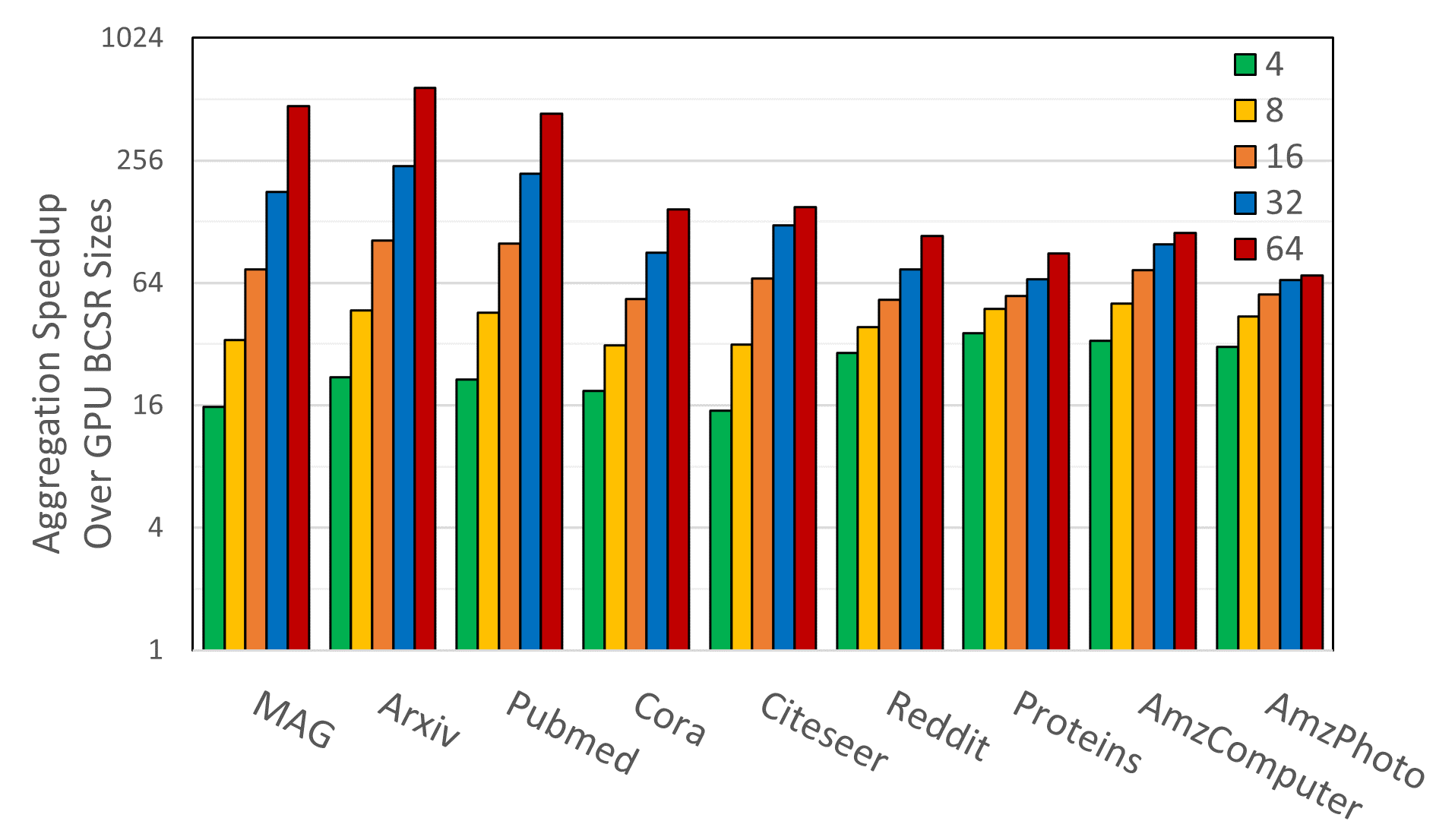}
    \caption{\red{Speedup of SCV-Z over various block sizes of BSCR.}}
    \label{fig:GPU}
\end{figure}

\begin{figure}[t]
    \centering
    \includegraphics[width=\linewidth]{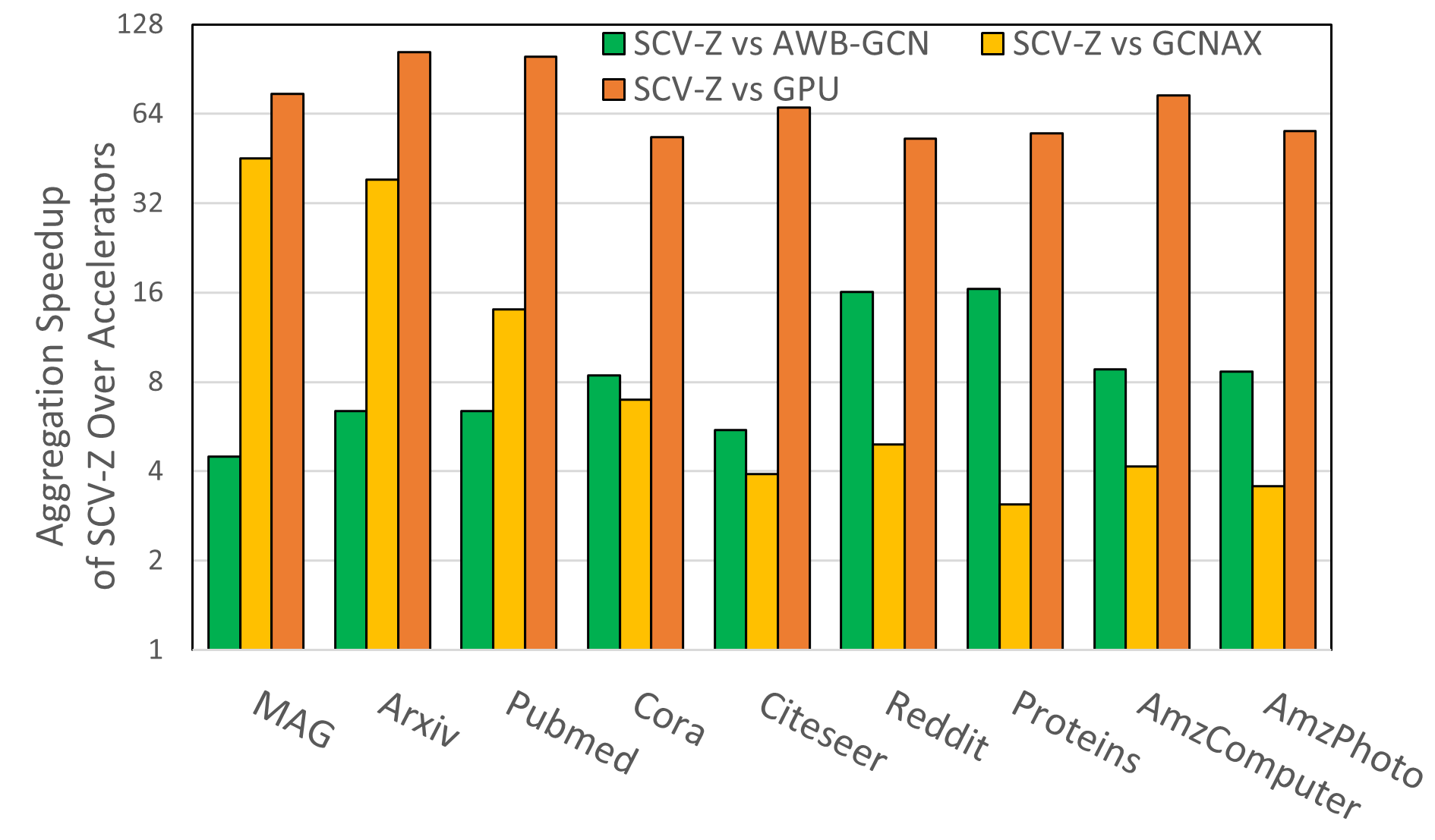}
    \caption{\red{Speedup of SCV-Z over GPU, AWB-GCN, and GCNAX for aggregation. GPU is modeled as BCSR with block size 16.} }
    \label{fig:accel-compare}
\end{figure}

\subsection{\red{Comparison With Previous Work}}
\red{We compare the performance of SCV-Z against current accelerators for the aggregation operation. The three accelerators we compare against are GPU\cite{GPU}, AWB-GCN\cite{AWB-GCN}, and GCNAX\cite{GCNAX}. 
GPUs use BSCR for computing the SpMM operation in aggregation. As explained in \cref{sec:spase}, this is not suited for the ultra-sparse nature of GNN aggregation, and the large memory overhead of storing blocks densely is worse for larger block sizes. However, GPUs are designed to perform more efficiently at larger block sizes and typically use sizes 16 or 32. We sweep block sizes from 4 to 64 and show the speedup of SCV-Z against the GPU in  \cref{fig:GPU}. 
AWB-GCN is an accelerator that uses CSC for storing the adjacency matrix and computes the outer product for SpMM processing. It performs efficient load balancing but suffers in partial sum memory accesses incurred by CSC and the outer product. 
GCNAX utilizes a well-known optimization for consecutive matrix multiplications, where tiles' processing order can be changed to reduce the number of memory accesses. Their system supports dynamically choosing to use this modified processing order based on the input matrices. However, their system is configured with non-columnar adjacency tiles, leading to inefficient feature matrix accesses. Because combination is orthogonal to the proposed method, we only compare it against their aggregation step.
The speedups across different datasets versus each accelerator are shown in \cref{fig:accel-compare}. We ensure that for each comparison, both the SCV-Z and baseline accelerators have an equivalent number of MACs and are allocated equal on-chip memories overall. For GPU, we use block size 16 BSCR as our comparison point. Across all datasets, we see geometric mean improvements of $68.5\times$, $8.2\times$, and $8.1\times$ for GPU, AWB-GCN, and GCNAX, respectively. Though we emulate the function of the other accelerators to the best of our ability, these speedups are estimates based on the processing order of the SpMM operation.}
\section{Related Work}
\label{sec:related_work}

Though many AI/ML hardware accelerators exist, few works target GNNs~\cite{AccelSurvey}. Most DNN accelerators have been targeted toward CNNs or transformers~\cite{permdnn,sigma,EIE,eyerissv2_19}, which do not translate well to the ultra sparse nature of graph neural networks. Additionally, graph-specific processors like~\cite{graph_proc} explore scatter-gather graph systems but do not explore them in the context of graph neural networks or graph tiling.

The unique challenges of GNNs have resulted in GNN-specific accelerators. One challenge is the extreme sparsity and uneven edge distribution within a graph. AWB-GCN~\cite{AWB-GCN} addresses this through vector queues and fine-grain workload balancing. This reduces under-utilization but necessitates strict mapping of PEs to rows of the adjacency. They do not explore flexible mapping strategies and use a CSC-based tiling strategy, which we have shown is suboptimal . Similar issues can be seen following the same mapping and tiling strategy for systolic arrays~\cite{versgnn}.

One approach to solving the irregularity problem for GNNs is through partitioning or reordering schemes~\cite{IGCN,RabitOrder,partitioning}. The aim is to separate the graph into highly connected clusters, increasing the efficiency of the fetch and compute. These subgraphs can either be parsed densely or combined with sparse multiplication operations to improve efficiency. While these approaches lead to significant benefits, they are orthogonal to the proposed improvements and can be combined with SCV to improve its efficiency further.

HyGCN~\cite{HyGCN} is another GNN accelerator that tackles the sparsity in aggregation through adaptive tiling with sliding and shrinking windows, reducing the number of redundant accesses. However, each tile can still contain multiple columns of non-zeros, must be dynamically computed, and their column-centric tiling strategy, like tiled CSC, is shown to be suboptimal. Other adaptive tiling approaches~\cite{rubik,GROW,Ast} have similar inefficiencies that our proposed method overcomes. Multipass approaches, like~\cite{GNNie}, avoid the effect of tiling ordering on the lower levels of the memory at the expense of additional computations and control. However, our experiments show that proper tiling and order, like the proposed method, can outperform multipass approaches on aggregate.

Cambricon-G~\cite{Cambricon-G} explores the data locality and tiling aspect of GNNs by envisioning the operations as cuboids. The use of cuboids allows for an efficient transfer of information between adjacent vector processing units. However, the fixed nature of the mapping and tiling has similar issues as blocked approaches like CSB and requires additional overhead for workload balancing and prefetching. In particular, due to the use of a tiled CSR approach for internal storage, the cuboid must have additional processing to ensure effective prefetching. SCV inherently stores this information locally, and the processing order is better optimized for workload balancing. While the work in~\cite{Omega} explored tiling strategies for GNNs, their analysis for aggregation was limited to CSR-based tiling which we have shown to be suboptimal. None of the existing tiling works have explored both the vector-based approach and locality-driven ordering of SCV-Z.

\section{Conclusion}
\label{sec:conclusion}

This paper proposes a novel sparse format, SCV/SCV-Z, that is designed to maximize GNN aggregation efficiency in inference. The proposed format maximizes parallelism while minimizing memory access during aggregation. The proposed method outperforms baseline sparse formats by an average factor of $6.51\times \sim 7.96\times$ over a variety of GNN datasets. \red{While the primary focus of the paper is the new sparse format for the aggregation step, any improvements for combination will be orthogonal to the proposed method and can be combined with the proposed method. We will explore further optimizations in future work.} Future work will be directed towards (i) adapting this work to enhance the training of GNNs, (ii) designing a unified hardware accelerator to support all GNN operations during training and inference, and (iii) \red{further co-design of the hardware and sparse format to improve performance}.

\bibliographystyle{IEEEtran}
\bibliography{references}

\begin{IEEEbiography}[{\includegraphics[width=1in,height=1.25in,clip,keepaspectratio]{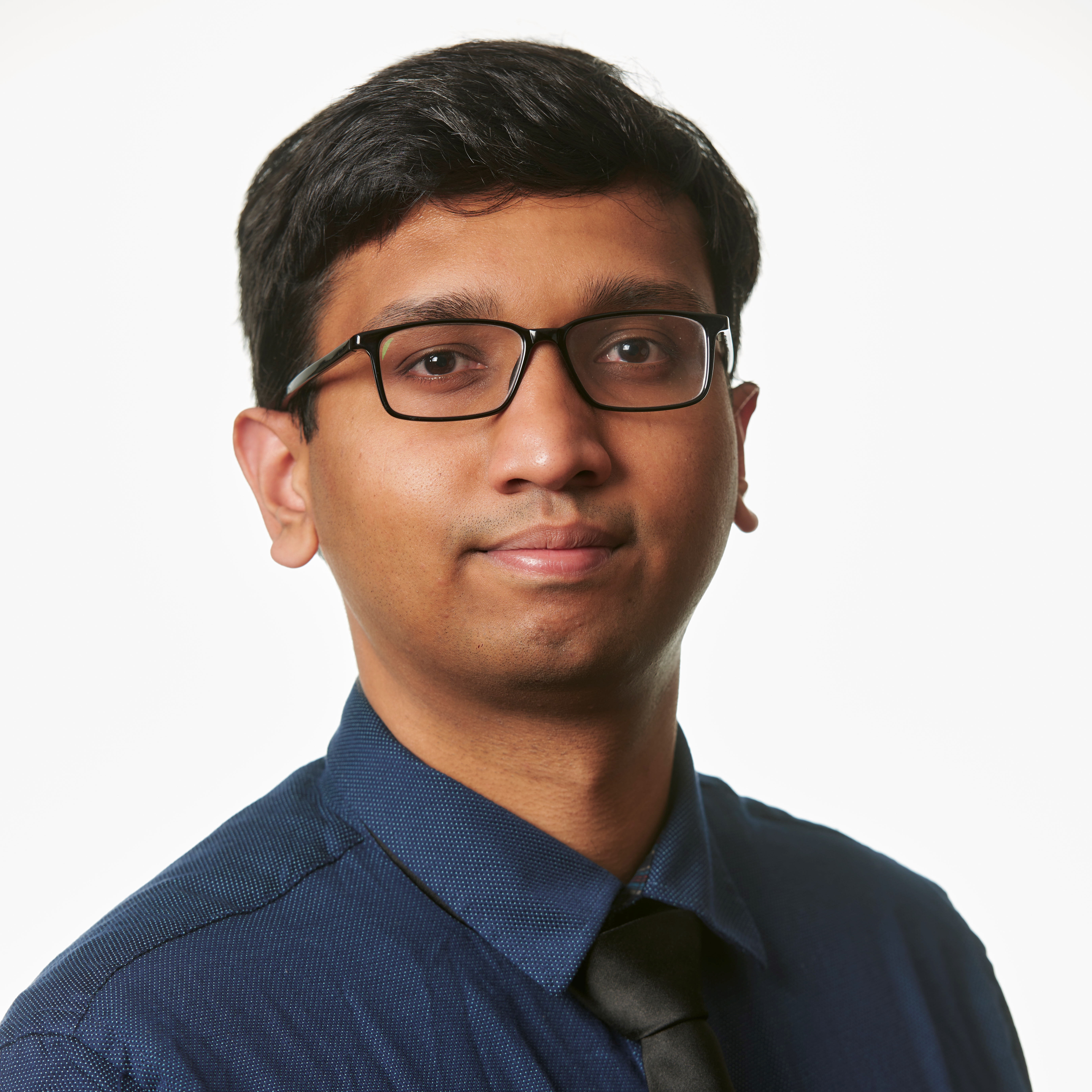}}]{Nanda K. Unnikrishnan} (Graduate student member, IEEE) is currently pursuing a Ph.D. degree in electrical engineering at the University of Minnesota, Minneapolis, USA.  He received his B.Tech in electronics and communication from the National Institute of Technology, Karnataka (NITK), Suratkal, India, in 2014 and his M.S. in electrical engineering from the University of Minnesota in 2018.  

He worked at SilabTech, India (now Synopsys) from 2014 to 2016 as a design verification engineer for mixed-signal designs. He has interned at Qualcomm Technologies Inc, San Diego in Summer 2017, with Intel Labs in Summer 2018 and Facebook in Summer 2021. His research interests lie in design of VLSI architectures for machine learning and deep learning systems.
\end{IEEEbiography}

\begin{IEEEbiography}[{\includegraphics[width=1in,height=1.25in,clip,keepaspectratio]{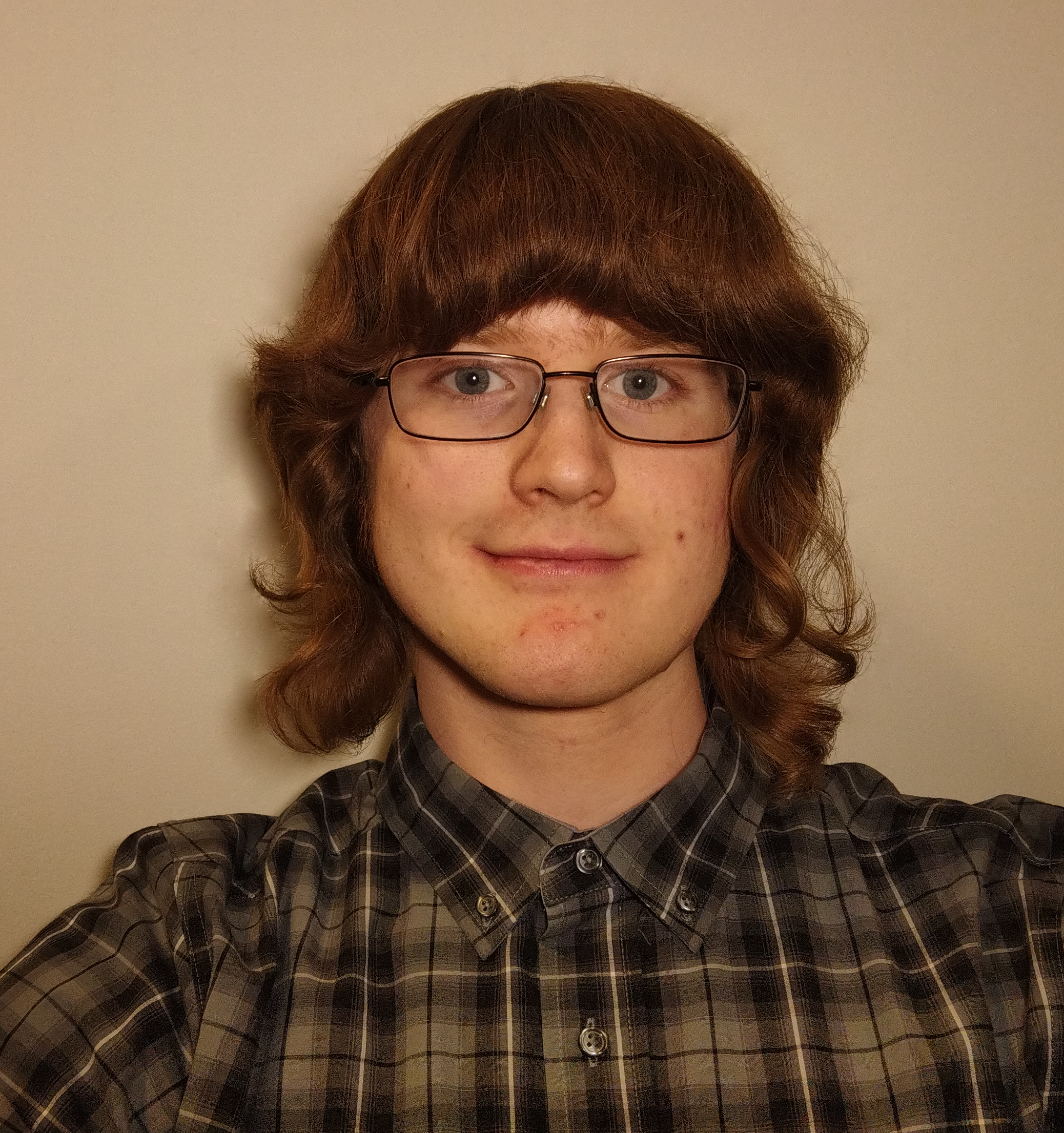}}]{Joe Gould} (Student member, IEEE) is currently pursuing an M.S. degree in electrical engineering at the University of Minnesota, Minneapolis, USA. 
He worked at Commscope, USA, from 2020 to 2021 under a co-op internship program as a design verification engineer for FPGA products. His interests are in the design of efficient architectures for accelerator and signal processing systems.

\end{IEEEbiography}

\begin{IEEEbiography}[{\includegraphics[width=1in,height=1.25in,clip,keepaspectratio]{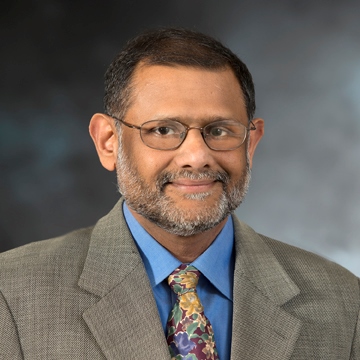}}]{Keshab K. Parhi} (Fellow, IEEE) is Distinguished McKnight University Professor and Erwin A. Kelen Chair in the Department of Electrical and Computer Engineering. He completed his Ph.D. in EECS at the University of California, Berkeley in 1988. He has published over 680 papers, is the inventor of 33 patents, and has authored the textbook VLSI Digital Signal Processing Systems (Wiley, 1999). His current research addresses VLSI architectures for machine learning, hardware security, data-driven neuroscience, and molecular/DNA computing. Dr. Parhi is the recipient of numerous awards, including the 2003 IEEE Kiyo Tomiyasu Technical Field Award, and the 2017 Mac Van Valkenburg award, and the 2012 Charles A. Desoer Technical Achievement award from the IEEE Circuits and Systems Society. He served as the Editor-in-Chief of the {\em IEEE Trans. Circuits and Systems, Part-I: Regular Papers} during 2004 and 2005. He is a Fellow of the ACM, AIMBE, AAAS, and NAI.
\end{IEEEbiography}

\end{document}